\documentclass[a4paper,11pt]{article}

\usepackage{jheppub}
\usepackage[utf8]{inputenc}
\usepackage{graphicx,cancel,float}
\usepackage{amssymb}
\usepackage{textcomp}
\usepackage{amsmath}
\usepackage{tabularx}
\usepackage{bm}
\usepackage{times}
\usepackage{color}
\usepackage{slashed}
\usepackage{multirow}
\usepackage{verbatim}
\usepackage{epsfig,epstopdf}
\usepackage{subfigure}
\allowdisplaybreaks[4]

\def\lsim{\mathrel{\raise.3ex\hbox{$<$\kern-.75em\lower1ex\hbox{$\sim$}}}}
\def\gsim{\mathrel{\raise.3ex\hbox{$>$\kern-.75em\lower1ex\hbox{$\sim$}}}}

\newcommand{\bew}{\begin{widetext}}
\newcommand{\enw}{\end{widetext}}
\newcommand{\bea}{\begin{eqnarray}}
\newcommand{\ena}{\end{eqnarray}}
\newcommand{\bes}{\begin{subequations}}
\newcommand{\ens}{\end{subequations}}

\definecolor{orange}{rgb}{1,0.5,0}

\subheader{\it \today}

\title{\Large{\bf Laser-assisted Light-by-Light Scattering in Born-Infeld and Axion-like Particle Theories}}

\author[a]{Kai Ma}
\emailAdd{kai@xauat.edu.cn}
\author[b]{Tong Li}
\emailAdd{litong@nankai.edu.cn}

\affiliation[a]{Faculty of Science, Xi'an University of Architecture and Technology, Xi'an, 710055, China}
\affiliation[b]{School of Physics, Nankai University, 94 Weijin Road, Tianjin 300071, China}

\abstract{
The precision measurements of well-known light-by-light reactions lead to important insights of nonlinear quantum electrodynamics (QED) vacuum polarization. The laser of an
intense electromagnetic field strength provides an essential tool for exploring nonlinear QED and new physics beyond Standard Model (SM) in the high-precision frontier. In this work, we propose to search for low-energy light-by-light scattering in the collision of a photon beam and a laser pulse of classical background field. We aim to investigate the impact of Born-Infeld (BI) and axion-like particle (ALP) theories on laser-assisted light-by-light scattering. We calculate the QED light-by-light scattering cross section using complete QED helicity amplitudes, and then combine them with the amplitudes in BI or ALP theory to evaluate the total cross section. The laser-assisted SM light-by-light scattering should be observable in future experiments with very moderate integrated luminosities. The sensitivity of laser-assisted light-by-light scattering to BI and ALP parameters is presented.
}

\begin{document}

\maketitle
\setcounter{page}{2}

\newpage

\section{Introduction}
\label{sec:Intro}

In quantum field theory, quantum effects induce non-trivial quantum vacuum state and vacuum polarization from the pair production of virtual particles. The vacuum polarization in quantum electrodynamics (QED) results in nonlinear QED phenomena and possible photon self-interaction. The nonlinear corrections to classical electrodynamics in vacuum are described by the low-energy effective Euler-Heisenberg (EH) Lagrangian~\cite{Heisenberg:1936nmg,Weisskopf:1936hya} (see a recent review Ref.~\cite{Dunne:2004nc}). A well-known consequence of the nonlinear QED effect is the light-by-light reaction~\cite{Karplus:1950zz}
\begin{eqnarray}
\gamma\gamma \to \gamma\gamma
\end{eqnarray}
through one-loop box diagram involving virtual electrons in QED. The precision measurement of this light-by-light reaction
provides crucial insights into QED vacuum polarization and its applications.
The light-by-light interaction was recently observed through real photon-photon scattering in relativistic heavy-ion
collision at the Large Hadron Collider (LHC)~\cite{dEnterria:2013zqi,ATLAS:2017fur,CMS:2018erd}. This high-energy measurement relies on fiducial cuts of the photon transverse energy $E_T>3~(2)$ GeV and the diphoton invariant mass $m_{\gamma\gamma}>6~(5)$ GeV at ATLAS~\cite{ATLAS:2017fur,ATLAS:2020hii} (CMS~\cite{CMS:2018erd,CMS:2024bnt}) experiments.

The EH Lagrangian can also generate distinct light-by-light processes of photons at low energies.
The low-energy processes originate from the scattering off a classical electromagnetic background, such as light-by-light
Delbr\"{u}ck scattering in the
electric field of a nucleus~\cite{Wilson:1953zz,Jarlskog:1973aui,Schumacher:1975kv,Akhmadaliev:1998zz}, and vacuum birefringence~\cite{Toll:1952rq,Mignani:2016fwz} or photon splitting~\cite{Akhmadaliev:2001ik} in a magnetic field. A recent proposal suggests the
detection of light-by-light scattering in the collision of a highly collimated broadband gamma-ray beam and an extreme ultraviolet (XUV) pulse~\cite{Sangal:2021qeg}.
Compared with high-energy collision experiments, the facilities with electromagnetic background field may enhance the probability of the assisted process and yield interesting phenomena.
In particular, the laser of an intense electromagnetic field strength provides an essential tool for exploring strong-field particle physics in the high-intensity frontier.
J. Schwinger showed that at field strength of $E=m_e^2/e\sim 1.32\times 10^{18}~{\rm V/m}$, the QED vacuum becomes unstable and can decay into a pair of electron and position~\cite{Schwinger:1951nm}. This so-called Schwinger field is of particular interest because it enables the study of non-perturbative QED effects.
Moreover, the presence of an intense laser field may change the
properties of elementary particles or induce processes that cannot occur in vacuum.
The measurement of the strong-field processes catalyzes the studies of nonlinear QED (see Greiner et al.'s textbook~\cite{Greiner:1992bv} and recent reviews in Refs.~\cite{Hartin:2018egj,Fedotov:2022ely}). Interestingly, the laser-assisted processes can also provide potentially useful search for new physics (NP) beyond the Standard Model (SM) such as proton decay~\cite{Wistisen:2020czu,Ouhammou:2022jys}, dark matter~\cite{Ma:2025axq,Fuchs:2024edo}, dark photon~\cite{Ma:2024ywm}, axion-like particle (ALP)~\cite{Dillon:2018ypt,King:2018qbq,Bai:2021gbm,Dillon:2018ouq,King:2019cpj,Beyer:2021mzq,Huang:2020lxo,Ma:2024ywm} and so on.

In this work, we propose to search for the signals of two famous NP benchmarks beyond pure QED using laser-assisted light-by-light scattering. In 1934, Born and Infeld proposed
a nonlinear generalization of the QED Lagrangian in terms of an unknown parameter $b$ with mass dimension 2~\cite{Born:1934gh}
\begin{eqnarray}
\mathcal{L}^{\rm BI}=b^2\Big(1-\sqrt{1+{1\over 2b^2}F_{\mu\nu}F^{\mu\nu}-{1\over 16b^4}(F_{\mu\nu}\tilde{F}^{\mu\nu})^2}\Big)\;,
\end{eqnarray}
where $F_{\mu\nu}$ is electromagnetic field strength tensor and $\tilde{F}_{\mu\nu}$ is its Hodge dual tensor.
By expanding the BI Lagrangian in powers of $1/b$,
higher order interactions, for instance, quartic photon vertices, etc., emerge.
Since the leading order contribution of the expansion always coincides with the free electromagnetic gauge term, there is no general constraint on the energy scale $\sqrt{b}$ above which the perturbative expansion is valid~\cite{Carley:2006zz,Davila:2013wba,Fouche:2016qqj}. Nevertheless, if $\sqrt{b}<m_e$, the BI model can lead to huge correction on the light-by-light scattering. Hence, $\sqrt{b} > m_e$ is a primary condition.
The observation of light-by-light scattering at the LHC placed a strong bound on the BI parameter $\sqrt{b} \gsim 100\,{\rm GeV}$~\cite{Ellis:2017edi}.
However, such constraint can only be valid for high invariant mass of photon pair $m_{\gamma\gamma}$, as the cross section grows very rapidly with $m_{\gamma\gamma}$ ($\sigma^{\rm BI} \propto m_{\gamma\gamma}^6$). In this work, we investigate the BI model at low-energy scale, {\it i.e.}, $m_{\gamma\gamma} \lsim m_e$, where only the primary condition $\sqrt{b} > m_e$ is assumed.

On the other hand,
the ALP, a pseudo-Nambu-Goldstone boson $a$ that accounts for the spontaneous breaking of a global $U(1)$ symmetry~\cite{Peccei:1977hh,Peccei:1977ur,Weinberg:1977ma,Wilczek:1977pj} (see a recent review Ref.~\cite{DiLuzio:2020wdo}), can also induce non-trivial light-by-light scattering with a rate depending on the mass of the ALP and the symmetry breaking scale.
In general, the ALP mass ($m_a$) and the symmetry breaking scale (or called decay constant $f_a$) may not be related~\cite{Dimopoulos:1979pp,Tye:1981zy,Zhitnitsky:1980tq,Dine:1981rt,Holdom:1982ex,Kaplan:1985dv,Srednicki:1985xd,Flynn:1987rs,Kamionkowski:1992mf,Berezhiani:2000gh,Hsu:2004mf,Hook:2014cda,Alonso-Alvarez:2018irt,Hook:2019qoh}. The mass range of ALP may span from sub-micro-eV~\cite{Kim:1979if,Shifman:1979if,Dine:1981rt,Zhitnitsky:1980tq,Turner:1989vc} to TeV scale and even beyond~\cite{Rubakov:1997vp,Fukuda:2015ana,Gherghetta:2016fhp,Dimopoulos:2016lvn,Chiang:2016eav,Gaillard:2018xgk,Gherghetta:2020ofz}. Thus, the search for ALPs requires rather different experimental strategies and facilities. The Lagrangian of photophilic ALP is
\begin{eqnarray}
\label{eq:L:ALP}
\mathcal{L}^{\rm ALP}_{\rm int}=-{1\over 4}g_{a\gamma\gamma}\,a \,F^{\mu\nu}\tilde{F}_{\mu\nu}\;.
\end{eqnarray}
This ALP-photon coupling can induce anomalous light-by-light scattering through a virtual ALP at tree level. There were quite a few proposals to search for ALP through photon-photon scattering at colliders~\cite{Knapen:2016moh,Baldenegro:2018hng,Inan:2020aal,Inan:2020kif,Harland-Lang:2022jwn,Balkin:2023gya,RebelloTeles:2023uig} and recent measurements based on photon-photon scattering in Pb+Pb~\cite{CMS:2018erd,ATLAS:2020hii,CMS:2024bnt} and proton-proton~\cite{ATLAS:2023zfc,TOTEM:2021zxa,TOTEM:2023ewz} collisions at the LHC.
In our proposal, a high-energy gamma-ray beam collides with an intense laser pulse and they then scatters to a pair of photons in these two frameworks
\begin{eqnarray}
\gamma + {\rm laser} \to \gamma + \gamma ~ {\rm in~BI~theory}~~{\rm or}~~ \gamma + {\rm laser} \to a\to \gamma + \gamma ~ {\rm in~ALP~theory}\;.
\end{eqnarray}
By taking into account the interference terms between NP and QED amplitudes, we calculate the total cross section of light-by-light scattering in these two NP extensions. The angular distributions of differential cross sections are then compared with that of pure QED. We will show the sensitivities of laser-induced light-by-light scattering to the BI parameter $b$ as well as the ALP-photon coupling $g_{a\gamma\gamma}$ and mass $m_a$.

This paper is organized as follows.
In Sec.~\ref{sec:LbyL}, we review the light-by-light scattering in terms of EH Lagrangian in QED.
We present the detailed calculations of the laser-induced light-by-light scattering in BI and ALP theories in Sec.~\ref{sec:BIALP}. The results of sensitivity reach for the BI parameter $b$ and the ALP-photon coupling $g_{a\gamma\gamma}$ are also shown in Sec.~\ref{sec:Sen}. Our conclusions are drawn in Sec.~\ref{sec:Con}.

\section{QED light-by-light scattering in vacuum or a classical background}
\label{sec:LbyL}

In this section we first review the QED light-by-light scattering $\gamma\gamma\to \gamma\gamma$ in vacuum. We then summarize the property of scattering in which an incoming photon is replaced by a classical background field.

Let's first consider the following real photon-photon scattering induced by the low-energy effective EH Lagrangian
\begin{eqnarray}
\gamma (p) + \gamma (k) \to \gamma (p') + \gamma (k')\;.
\end{eqnarray}
For the case with center-of-mass (c.m.) energy much less than electron mass, the quartic term in EH Lagrangian is
\begin{eqnarray}
\mathcal{L}_4^{\rm EH}={2\alpha^2\over 45m_e^4}\Big[ {7\over 4}{\rm tr}(F^4)- {5\over 8}({\rm tr}(F^2))^2\Big]\;,
\end{eqnarray}
where the identity $(F\cdot \tilde{F})^2=4{\rm tr}(F^4)-2({\rm tr}(F^2))^2$ is used.
The EH amplitude is then given by
\begin{eqnarray}
\mathcal{M}^{\rm EH}&=&{2\alpha^2\over 45m_e^4}\Big[7{\rm tr}(F_p F_{k} F_{p'} F_{k'}+F_{p'} F_p F_{k} F_{k'}+F_{k} F_{p'} F_p F_{k'}+F_{p'}F_k F_p F_{k'}+F_p F_{p'} F_k F_{k'}\nonumber\\
&&+F_{k} F_p F_{p'} F_{k'})-5\Big({\rm tr}(F_p F_k){\rm tr}(F_{p'}F_{k'})+{\rm tr}(F_p F_{p'}){\rm tr}(F_{k}F_{k'})+{\rm tr}(F_p F_{k'}){\rm tr}(F_{k}F_{p'})\Big)\Big]\;,\nonumber\\
\end{eqnarray}
where $F_{\ell\mu\nu}=\ell_\mu \varepsilon_\nu - \ell_\nu \varepsilon_\mu$ with $\varepsilon$
being the polarization of real photon and 4-momentum $\ell=p,k,p',k'$.
After summing over the polarizations and averaging those of initial states, we reproduce the squared EH amplitude~\cite{Davila:2013wba}
\begin{eqnarray}
\overline{|\mathcal{M}^{\rm EH}|^2}&=&{4\alpha^4\over 45^2 m_e^8} \Big[314\Big((p\cdot k')^2(k\cdot p')^2+(p\cdot p')^2(k\cdot k')^2+(p\cdot k)^2(p'\cdot k')^2\Big)\nonumber\\
&&-72\Big(p\cdot k' k\cdot p' p\cdot p' k\cdot k'+p\cdot k' k\cdot p' p\cdot k p'\cdot k' + p\cdot k p\cdot p' k\cdot k' p'\cdot k'\Big) \Big]\nonumber\\
&=&{8\cdot 139\alpha^4\over 45^2 m_e^8}\Big((p\cdot k')^2(k\cdot p')^2+(p\cdot p')^2(k\cdot k')^2+(p\cdot k)^2(p'\cdot k')^2\Big)\;,
\end{eqnarray}
where the relations of Mandelstam variables $s+t+u=0$ and $s^4+t^4+u^4=2(s^2t^2+s^2u^2+t^2u^2)$ are used with $s=(p+k)^2$, $t=(p-p')^2$ and $u=(p-k')^2$. After including an additional $1/2$ for identical outgoing photons, the unpolarized cross-section is
\begin{eqnarray}
{d\sigma^{\rm EH}\over d\Omega}={1\over 2}{1\over 64 \omega^2_{\rm CM} (2\pi)^2} \overline{|\mathcal{M}^{\rm EH}|^2} = {1\over 2}{139 \alpha^2 r_e^2\over 32400\pi^2} \Big({\omega_{\rm CM}\over m_e}\Big)^6 (3+\cos^2\theta)^2\;,
\end{eqnarray}
where the classical electron radius is defined as $r_e=\alpha/m_e$,
$\omega_{\rm CM}=\sqrt{s}/2$ and $\theta$ denote the beam energy and scattering angle in the c.m. frame, respectively.
The total cross-section is given by
\begin{eqnarray}
\sigma^{\rm EH}={973 \alpha^2 r_e^2\over 10125\pi} \Big({\omega_{\rm CM}\over m_e}\Big)^6\;.
\end{eqnarray}
This cross-section agrees with that in Refs.~\cite{Davila:2013wba,Heinzl:2024cia}.

One notes that the above calculation is only valid in low-energy regime $\sqrt{s}\lsim m_e$.
To obtain reliable results even in the relativistic limit, we must incorporate the complete QED corrections to the electron loop contribution in light-by-light scattering.
The complete unpolarized differential cross section is given by the one-loop helicity amplitudes of QED as~\cite{Jikia:1993tc,Gounaris:1998qk,Gounaris:1999gh,Bern:2001dg,Inan:2020aal}
\begin{eqnarray}
{d\sigma^{\rm QED}\over d\Omega}&=&{1\over 2}{1\over 64\pi^2 s} \overline{\sum |\mathcal{M}^{\rm QED}|^2}\nonumber\\
&=&{1\over 256\pi^2 s}\Big(|\mathcal{M}^{\rm QED}_{++++}|^2+4|\mathcal{M}^{\rm QED}_{-+++}|^2+|\mathcal{M}^{\rm QED}_{--++}|^2+|\mathcal{M}^{\rm QED}_{+--+}|^2+|\mathcal{M}^{\rm QED}_{+-+-}|^2\Big)\;,
\label{eq:EHxsec}
\end{eqnarray}
where the complete results of helicity amplitude are shown in Appendix A of Ref.~\cite{Gounaris:1999gh}.
Here we only consider the contribution from electron, and hence the following results are only valid up to the mass scale of muon lepton. For completeness and clarity, we show the related formula explicitly bellow
\begin{eqnarray}
\label{eq:amp:QED:12}
\mathcal{M}^{\rm QED}_{++++}(s,t,u)&=&\alpha^2 e^4 \Big[-8 + 8\Big(1+{2u\over s}\Big)B_0(u)+ 8\Big(1+{2t\over s}\Big)B_0(t)\nonumber\\
&&-8\Big({t^2+u^2\over s^2}-{4m_e^2\over s}\Big)(tC_0(t)+uC_0(u))\nonumber\\
&&+8m_e^2(s-2m_e^2)(D_0(s,t)+D_0(s,u))\nonumber\\
&&-4\Big(4m_e^4-(2sm_e^2+tu){t^2+u^2\over s^2}+{4m_e^2tu\over s}\Big)D_0(t,u)\Big]\;,\\
\mathcal{M}^{\rm QED}_{-+++}(s,t,u)&=&-{2\over 3}\alpha^2 e^4 \Big[-12+24m_e^4 (D_0(s,t)+D_0(s,u)+D_0(t,u))\nonumber\\
&&+12m_e^2 stu \Big({D_0(s,t)\over u^2}+{D_0(s,u)\over t^2}+{D_0(t,u)\over s^2}\Big)\nonumber\\
&&-24m_e^2\Big({1\over s}+{1\over t}+{1\over u}\Big)(tC_0(t)+uC_0(u)+sC_0(s))\Big]\;,\\
\mathcal{M}^{\rm QED}_{--++}(s,t,u)&=&-{2\over 3}\alpha^2 e^4 \Big[-12 + 24m_e^4 (D_0(s,t)+D_0(s,u)+D_0(t,u))\Big]\;,\\
\mathcal{M}^{\rm QED}_{+--+}(s,t,u)&=&\mathcal{M}^{\rm QED}_{++++}(t,s,u)=\mathcal{M}^{\rm QED}_{++++}(t,u,s)\;,\\
\label{eq:amp:QED:18}
\mathcal{M}^{\rm QED}_{+-+-}(s,t,u)&=&\mathcal{M}^{\rm QED}_{++++}(u,t,s)\;.
\end{eqnarray}
The one-loop Passarino-Veltman functions~\cite{Passarino:1978jh} are defined as
\begin{eqnarray}
&&B_0(s)\equiv B_0(s;m_e,m_e)\;,~~C_0(s)\equiv C_0(0,0,s;m_e,m_e,m_e)\;,\\
&&D_0(s,t)\equiv D_0(0,0,0,0,s,t;m_e,m_e,m_e,m_e)\;.
\end{eqnarray}

\begin{figure}[tb!]
\centering
\includegraphics[width=0.7\textwidth]{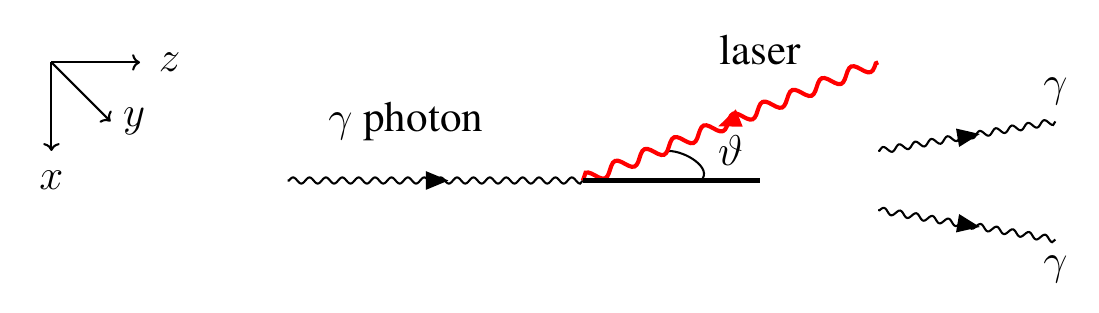}
\caption{Schematic diagram for the laser-assisted light-by-light scattering.
}
\label{fig:Feynman}
\end{figure}

Next we review the QED light-by-light scattering in a classical background. Fig.~\ref{fig:Feynman} shows the schematic diagram for the laser-assisted light-by-light scattering. It turns out that one of the incoming photon in the above $2\to 2$ scattering is replaced by an intense laser pulse.
Since the energy of gamma-ray is much larger than that of the laser photon, the outgoing photons are nearly collinear with the incoming gamma-ray. As a result, it is nearly impossible to resolve the two outgoing photons. Ref.~\cite{Sangal:2021qeg} showed that asymmetric beam topology can be helpful to identify the two outgoing photons. Despite that the angular distribution is still challenging to measure, we use this asymmetric topology for our analysis. In this case, the c.m. energy is given by
\begin{equation}
\sqrt{s}=2\sqrt{\omega_\gamma \omega_L} \cos\frac{\vartheta}{2}\;,
\label{eq:sqrts}
\end{equation}
where $\omega_\gamma$ ($\omega_L$) is the incoming photon beam energy (laser beam energy) and $\vartheta$ is the angle between the photon and laser beams. The details of Lorentz transformation from c.m. frame to laboratory (Lab) frame are collected in Appendix~\ref{app:Lab}.
Recently, Ref.~\cite{Heinzl:2024cia} derived the probability and cross-section of this $1\to 2$ process~\footnote{The numbers represent the real photons in initial or final state of scattering process.}. We review their evaluation here.
For the classical background field, one defines a field-dependent form factor
\begin{eqnarray}
F_q^{\mu\nu}\equiv \chi^{\mu\nu}(q)=\chi^{(1)}(q)\epsilon^{\mu\nu}=\int d^4k \chi^{\mu\nu}(k)\delta^4(k-q)\;,
\end{eqnarray}
where $q=p'+k'-p$, and $\epsilon^{\mu\nu}=n^\mu \epsilon^\nu-n^\nu \epsilon^\mu$ with $n^\mu$ and $\epsilon^\mu$ being a light-like vector and the polarization of classical field, respectively. Then, the $T$ amplitude of this $1\to 2$ process is given by
\begin{eqnarray}
T_{fi}^{1\to 2}=\sqrt{2V\over \omega_k} \chi^{(1)}(q) \mathcal{M}^{\rm EH}(p,k,p',k')\;,
\end{eqnarray}
where $V$ denotes the normalization volume.
The probability of $1\to 2$ scattering is
\begin{eqnarray}
P^{1\to 2}&=&V^2\int {d^3 p' d^3 k'\over (2\pi)^6} |T^{1\to 2}_{fi}|^2\nonumber\\
&=&\int {d^3 k\over (2\pi)^3 2\omega_k} 2 |\overline{A}(k)|^2 P^{2\to 2}(k)\;,
\end{eqnarray}
where $\overline{A}$ is the amplitude of the background potential which is related to the form factor $\chi^{\mu\nu}$ through Fourier transformation.
The differential cross-section reads
\begin{eqnarray}
{d\sigma^{1\to 2}\over d\Omega}={1\over J}{V\over T}{dP^{1\to 2}\over d\Omega}\;,~~J=\int{d^3k\over (2\pi)^3 2\omega_k} 2 |\overline{A}(k)|^2\;,
\end{eqnarray}
where $J$ is the incoming number flux.
Finally, the differential scattering cross-section becomes
\begin{eqnarray}
{d\sigma^{1\to 2}\over d\Omega}={d\sigma^{2\to 2}\over d\Omega}(k,\varepsilon_k\to \epsilon_k)\;.
\label{eq:12xsec}
\end{eqnarray}
It turns out that the cross-section of $1\to 2$ scattering can be obtained by replacing one of
the incoming photon (denoted by momentum $k$) polarizations of the $2\to 2$ process by the fixed polarization of the classical background. In terms of a laser pulse with linear polarization, the cross-sections of $2\to 2$ process in vacuum and laser-assisted $1\to 2$ process have no difference.

\section{Laser-assisted light-by-light scattering in BI and ALP theories}
\label{sec:BIALP}

In this section, we perform calculations of laser-assisted light-by-light scattering in BI and ALP theories. We generally take the c.m. energy $\sqrt{s}$ as a facility input by combining the incoming beam energies and angle in Eq.~(\ref{eq:sqrts}). As a benchmark of experimental setups, we will also show the results for a collision of photon beams from the proposal in Ref.~\cite{Sangal:2021qeg}. In Ref.~\cite{Sangal:2021qeg} they consider gamma photon generation from the Compton scattering of an electron beam with 13 GeV mean energy and a 30 TW linearly polarized intense laser pulse. Similar electron beam and laser parameters are available at existing facilities such as FACET-II~\cite{Yakimenko:2019sya} and the European X-Ray Free-Electron Laser (XFEL)~\cite{Altarelli:2006zza}. A majority of photons passing through the collimator have energy of $\omega_\gamma=3.25$ GeV. We then require the gamma-ray to collide with an intense laser beam of green light with $\omega_L=2.35$ eV similar to SLAC experiments~\cite{Burke:1997ew,Bamber:1999zt} and a crossing angle $\vartheta=10^\circ$ which corresponds to c.m. energy $\sqrt{s}\approx 0.34m_e$. Our choice of a green laser is more conservative than the XUV pulse in Ref.~\cite{Sangal:2021qeg} delivered by FELs such as FLASH (photon energy range of $20-360$ eV)~\cite{Ackermann:2007zzd} or FERMI (photon energy range of $20-310$ eV)~\cite{PhysRevX.7.021043} which can cover the range of $\sqrt{s}> m_e$ discussed below.

\subsection{BI theory}

The leading order non-trivial contribution from the effective Born-Infeld Lagrangian is given by~\cite{Davila:2013wba,Rebhan:2017zdx}
\begin{eqnarray}
\mathcal{L}_4^{\rm BI}={1\over 32b^2}\Big[ 4{\rm tr}(F^4)- ({\rm tr}(F^2))^2\Big]\;.
\end{eqnarray}
As mentioned before, we assume that the above effective Lagrangian is valid only for $\sqrt{b}> m_e$.
Similar to the low-energy EH theory, the BI amplitude then becomes
\begin{eqnarray}
\mathcal{M}^{\rm BI}&=&{1\over 32b^2}\Big[16{\rm tr}(F_p F_{k} F_{p'} F_{k'}+F_{p'} F_p F_{k} F_{k'}+F_{k} F_{p'} F_p F_{k'}+F_{p'}F_k F_p F_{k'}+F_p F_{p'} F_k F_{k'}\nonumber\\
&&+F_{k} F_p F_{p'} F_{k'})-8\Big({\rm tr}(F_p F_k){\rm tr}(F_{p'}F_{k'})+{\rm tr}(F_p F_{p'}){\rm tr}(F_{k}F_{k'})+{\rm tr}(F_p F_{k'}){\rm tr}(F_{k}F_{p'})\Big)\Big]\;.\nonumber\\
\end{eqnarray}
Since both the QED and BI correction contribute the light-by-light scattering process, the total amplitude should be a coherent sum of the two contributions.
Hence, the non-trivial interference effect, which is usually considerable larger than the pure BI contribution, is expected. We will study this in details below.
At low energy with $\sqrt{s} \lsim m_e$, the total amplitude can be given as
\begin{eqnarray}
\mathcal{M}^{\rm EH+BI}=\mathcal{M}^{\rm EH} + \mathcal{M}^{\rm BI}\;.
\end{eqnarray}
The corresponding total cross section can be calculated analytically in a straightforward way, and is given as
\begin{eqnarray}
{d\sigma^{\rm EH+BI}\over d\cos\theta}
=
{ 1 \over 64  \pi} \frac{ \omega_{\rm CM}^6 }{ m_e^8 }
\left( \frac{556 \alpha^4 }{2025}  + \frac{1980 \alpha^2 }{2025}   \frac{m_e^4}{b^2} + \frac{m_e^8}{b^4} \right)
(3+\cos^2\theta)^2\;.
\end{eqnarray}
It agrees with the result in Ref.~\cite{Davila:2013wba}.
One can see that the interference contribution ($\propto b^{-2}$) is always larger than the pure BI contribution ($\propto b^{-4}$) as long as $\sqrt{b}>\sqrt[4]{2025/1980} \, m_e/\alpha \approx \alpha^{-1}m_e$. At higher energy scale with $\sqrt{s} \gsim m_e$, the full QED results given in equations from Eq.\eqref{eq:amp:QED:12} to Eq.\eqref{eq:amp:QED:18} have to be used.
In helicity basis, the BI amplitudes are~\cite{Rebhan:2017zdx}~\footnote{The amplitudes in Ref.~\cite{Rebhan:2017zdx} should be multiplied by an imaginary unit ``$i$'' to match our convention.}
\begin{eqnarray}
\mathcal{M}_{++++}^{\rm BI}&=&0\;,\\
\mathcal{M}_{-+++}^{\rm BI}&=&0\;,\\
\mathcal{M}_{--++}^{\rm BI}&=&{1\over 2b^2}s^2\;,\\
\mathcal{M}_{+--+}^{\rm BI}&=&{1\over 2b^2}u^2\;,\\
\mathcal{M}_{+-+-}^{\rm BI}&=&{1\over 2b^2}t^2\;.
\end{eqnarray}
We can then combine them with the complete QED helicity amplitudes in Eq.~(\ref{eq:EHxsec}) and evaluate the total cross section valid in high-energy regime.

\begin{figure}[htb!]
\begin{center}
\includegraphics[width=0.48\textwidth]{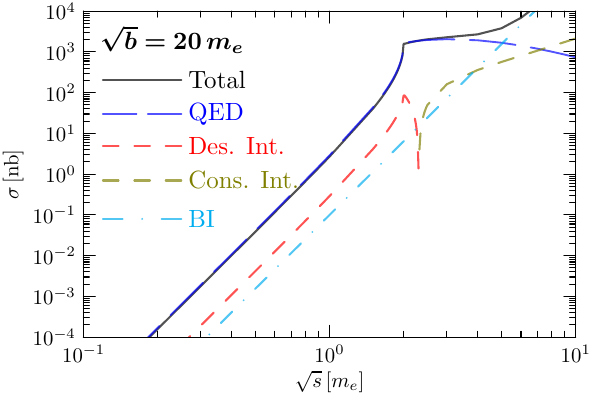}
\hfill
\includegraphics[width=0.48\textwidth]{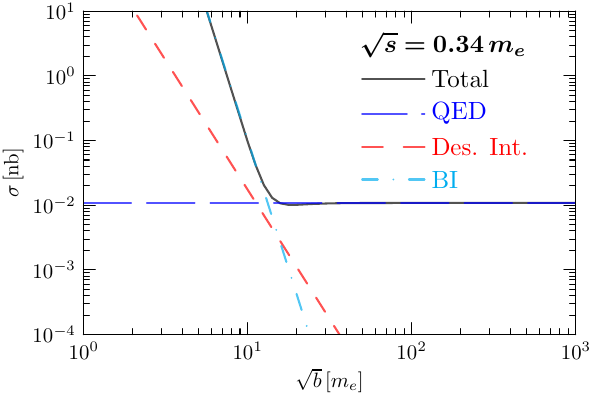}
\caption{The light-by-light scattering cross sections in BI theory as a function of c.m. energy $\sqrt{s}$ (left) and $\sqrt{b}$ (right). They include QED contribution (blue long-dashed line), BI contribution (cyan dash-dotted line), destructive interference (red dashed line), constructive interference (olive dashed line) and total cross section (black solid line). For illustration, we fix $\sqrt{b}=20m_e$ ($\sqrt{s}=0.34m_e$) on the left (right) panel.
}
\label{fig:xsBI}
\end{center}
\end{figure}
In Fig.~\ref{fig:xsBI}, we display the light-by-light scattering cross sections in BI theory as a function of c.m. energy $\sqrt{s}$ (left) and $\sqrt{b}$ (right).
We fix $\sqrt{b}=20m_e$ ($\sqrt{s}=0.34m_e$) on the left (right) panel for illustration. The QED cross section increases with rising $\sqrt{s}$, reaches its maximum $\sim 2\times 10^3~{\rm nb}$ at $\sqrt{s}=2m_e$, and subsequently decreases. The BI contribution grows (declines) with increasing $\sqrt{s}$ ($\sqrt{b}$). The interference term switches from destructive to constructive for the c.m. energy just above the threshold at $\sqrt{s}=2m_e$. The total cross section thus exhibits a knee-like feature in its $\sqrt{s}$ dependence. As $\sqrt{b}$ increases, the total cross section decreases and attains a minimum value at some critical $\sqrt{b}$ because the absolute value of negative interference term exceeds the pure BI contribution.
The cross section becomes a constant for $\sqrt{b}\gtrsim 15 m_e$ and the SM contribution completely dominates over the signal.

\begin{figure}[htb!]
\begin{center}
\includegraphics[width=0.48\textwidth]{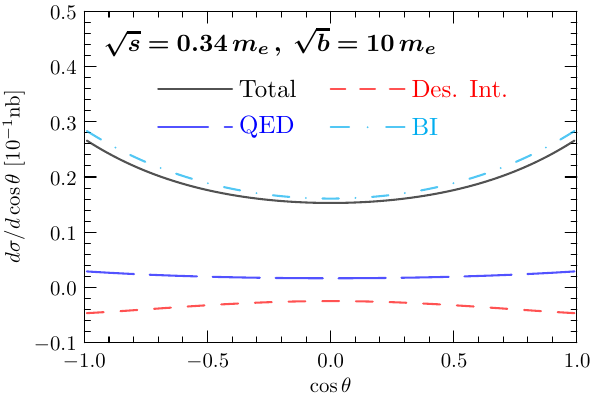}
\;
\includegraphics[width=0.48\textwidth]{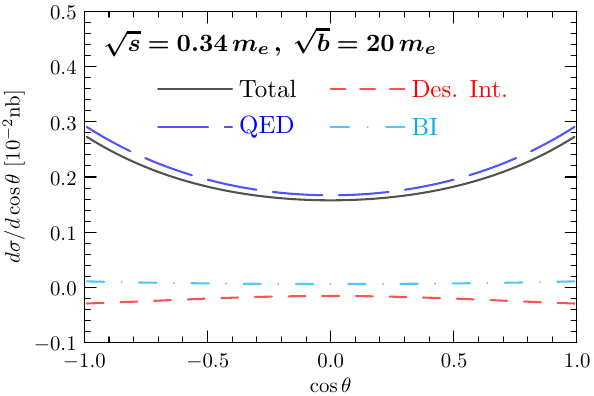}
\caption{The differential cross sections $d\sigma/d\cos\theta$ in BI theory with $\sqrt{s}=0.34m_e$ and $\sqrt{b}=10m_e$ (left) or $\sqrt{b}=20m_e$ (right). They include QED contribution (blue long-dashed line), BI contribution (cyan dash-dotted line), destructive interference (red dashed line), and total cross section (black solid line).
}
\label{fig:xcosBI}
\end{center}
\end{figure}
Fig.~\ref{fig:xcosBI} shows the differential cross section $d\sigma/d\cos\theta$ with $\sqrt{s}=0.34m_e$ and $\sqrt{b}=10m_e$ (left) or $\sqrt{b}=20m_e$ (right). One can see that larger $\sqrt{b}$ induces smaller BI contribution (cyan dash-dotted line) and destructive interference (red dashed line) in this case. As a result, the total cross section (black solid line) is a bit smaller than BI contribution for $\sqrt{b}=10m_e$ or pure QED contribution (blue long-dashed line) for $\sqrt{b}=20m_e$.

\begin{figure}[ht]
\begin{center}
\includegraphics[width=0.48\textwidth]{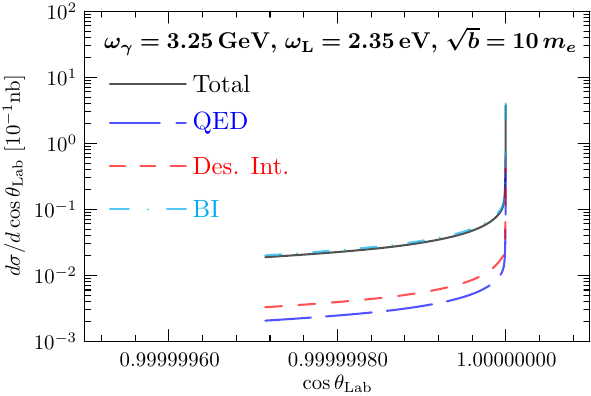}
\;
\includegraphics[width=0.48\textwidth]{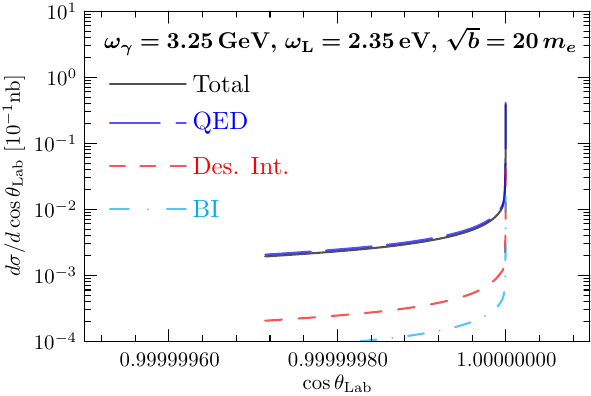}
\caption{The differential cross sections $d\sigma/d\cos\theta_{\rm Lab}$ in BI theory with $\omega_\gamma =3.25 {\rm GeV}$, $\omega_L =2.35 {\rm eV}$, $\vartheta=10^\circ$ and $\sqrt{b}=10m_e$ (left) or $\sqrt{b}=20m_e$ (right). They include QED contribution (blue long-dashed line), BI contribution (cyan dash-dotted line), destructive interference (red dashed line), and the total contribution (black solid line).
}
\label{fig:xcoslab:BI:xs}
\end{center}
\end{figure}
As we have mentioned, in practical scattering, the energies of the incoming two photons are highly unbalanced. As a result,
the outgoing photons are nearly collinear with the incoming gamma-ray. It was shown that
the asymmetric beam topology can help to identify the two outgoing photons~\cite{Sangal:2021qeg}.
Fig.~\ref{fig:xcoslab:BI:xs} shows the differential cross sections $d\sigma/d\cos\theta_{\rm Lab}$ with incoming photon energy $\omega_\gamma =3.25 \,{\rm GeV}$, $\omega_L =2.35 \,{\rm eV}$ (green laser) and an intersection angle $\vartheta=10^\circ$,
which corresponds to
a c.m. energy of $\sqrt{s} \approx 0.34m_e$.
The left and right panels stand for the model parameters $\sqrt{b}=10m_e$ and $\sqrt{b}=20m_e$, respectively. One can clearly see that in either case, most of the outgoing photons are along the direction of the incoming gamma-ray. One can also see the difference between the QED background and BI correction in the forward region.
In order to resolve such a small difference,
the uncertainty of the polar angle measurement should reach $10^{-4} \sim 10^{-3}$. It is challenging, but is achievable for a spatial resolution $\sim 10\,\mu{\rm m}$ and a detector placed $0.1~{\rm m}$ away from the interaction point.~\footnote{For a detector placed 0.1 m away from the interaction point with a 10-$\mu$m spatial resolution of photon reconstruction,
we obtained
$\sin\Delta\theta = 10\,\mu {\rm m}/0.1{\,\rm m} = 10^{-4}$,
where $\Delta\theta$ is the angle spanned by the 10-$\mu$m spatial resolution. For such a small angle we have $\Delta\theta \approx \sin\Delta\theta = 10^{-4}$.
This is roughly the minimal condition to resolve the scattered photons.
Detectors with improved spatial resolution and positioned farther from the interaction point can provide better angular resolution and thus higher detection efficiency.}

\subsection{ALP theory}

In general, the ALP can couple to all SM particles. Here we only consider its interaction with the photon, i.e., a photophilic ALP. The relevant Lagrangian is given by the Eq.~\eqref{eq:L:ALP}.
The Feynman rule between the ALP and two SM gauge bosons ($V_1$ and $V_2$) turns out to be~\footnote{The complete Feynman rules from the bosonic ALP effective Lagrangian can be found in the Appendix B of Ref.~\cite{Brivio:2017ije}.}
\begin{eqnarray}
-i g_{a V_1 V_2}~p_{V_1 \alpha} p_{V_2 \beta}~\epsilon^{\mu\nu\alpha\beta} \;,
\label{eq:tensor}
\end{eqnarray}
with the momenta ($p_{V_1}$, $p_{V_2}$) flowing inwards in the vertices.
For a photophilic ALP, there are two processes to which the ALP can contribute at leading order:
(1) the $\gamma\gamma \to a \to \gamma\gamma$ process
received non-trivial contribution from virtual ALP; (2) resonant production $\gamma\gamma \to a$ with $\sqrt{s}\approx m_a$; (3) on-shell production of the ALP from the external photon legs via the process $\gamma\gamma \to \gamma\gamma a$.
For the case (3), the ALP can be emitted from either one of the four external photon legs, through $\gamma^\ast \to \gamma a$ in final state emission or $\gamma \to \gamma^\ast a$ in initial state emission.
However, the case (3) process is irrelevant for two reasons. First, there is no interference contribution between QED light-by-light process and this on-shell ALP production.
Second, this process is suppressed by a factor of
$s^3/m_e^6$ as the pole structure of the off-shell photon $\gamma^\ast$ is essentially canceled by the momentum dependence in $a\gamma\gamma$ vertex.
The resultant cross section is very small in low-energy region with $\sqrt{s} < m_e$. 
Hence, next we only consider the cases (1) and (2).

The ALP contributes to light-by-light scattering through $s$, $t$ and $u$ channels.
The ALP amplitude then becomes
\begin{eqnarray}
\mathcal{M}^{\rm ALP}&=&g_{a\gamma\gamma}^2 \varepsilon_p^\mu \varepsilon_k^\nu \varepsilon_{p'}^{\ast\alpha}\varepsilon_{k'}^{\ast\beta} \Big[(-i)\epsilon_{\mu\nu\sigma\rho}p^\sigma k^\rho (-i)\epsilon_{\alpha\beta\delta\gamma}p^{\prime \delta} k^{\prime \gamma} {1\over (p+k)^2-m_a^2+im_a\Gamma_a}\nonumber\\
&&+i\epsilon_{\mu\alpha\sigma\rho}p^\sigma p^{\prime\rho} i\epsilon_{\nu\beta\delta\gamma}k^{\delta} k^{\prime \gamma} {1\over (p-p')^2-m_a^2}\nonumber\\
&&+i\epsilon_{\mu\beta\sigma\rho}p^\sigma k^{\prime\rho} i\epsilon_{\alpha\nu\delta\gamma}p^{\prime \delta} k^{\gamma} {1\over (p-k')^2-m_a^2}\Big]\;,
\end{eqnarray}
with the total decay width given by
\begin{eqnarray}
\Gamma_a\equiv \Gamma(a\to \gamma\gamma)={g_{a\gamma\gamma}^2 m_a^3\over 64\pi}\;.
\label{eq:width}
\end{eqnarray}
After ignoring $\Gamma_a$, we have
\begin{eqnarray}
\overline{|\mathcal{M}^{\rm ALP}|^2} &=& {g_{a\gamma\gamma}^4\over 16} \Big[{s^4\over (s-m_a^2)^2}+{t^4\over (t-m_a^2)^2}+{u^4\over (u-m_a^2)^2} \nonumber\\
&&+{s^2 t^2\over (s-m_a^2)(t-m_a^2)}+{s^2 u^2\over (s-m_a^2)(u-m_a^2)}+{t^2 u^2\over (t-m_a^2)(u-m_a^2)}\Big]\;.
\end{eqnarray}
Note that we ignore $\Gamma_a$ here only for simply showing the matrix element square. In our realistic calculation, we use the following helicity amplitudes in which $\Gamma_a$ is kept based on the above decay width in Eq.~(\ref{eq:width}).
The helicity amplitudes for ALP are~\cite{Baldenegro:2018hng,Inan:2020aal,Inan:2020kif}
\begin{eqnarray}
\mathcal{M}_{++++}^{\rm ALP}&=&-{g_{a\gamma\gamma}^2\over 4}{s^2 (s-m_a^2)\over (s-m_a^2)^2+m_a^2 \Gamma_a^2} + i {g_{a\gamma\gamma}^2\over 4} {s^2 m_a \Gamma_a\over (s-m_a^2)^2+m_a^2 \Gamma_a^2}\;,\\
\mathcal{M}_{-+++}^{\rm ALP}&=&0\;,\\
\mathcal{M}_{--++}^{\rm ALP}&=&{g_{a\gamma\gamma}^2\over 4}\Big[{s^2 (s-m_a^2)\over (s-m_a^2)^2+m_a^2 \Gamma_a^2} +{t^2\over t-m_a^2}+{u^2\over u-m_a^2}\Big]\nonumber\\
&&-i{g_{a\gamma\gamma}^2\over 4}{s^2 m_a \Gamma_a\over (s-m_a^2)^2+m_a^2 \Gamma_a^2}\;,\\
\mathcal{M}_{+--+}^{\rm ALP}&=&-{g_{a\gamma\gamma}^2\over 4}{t^2\over t-m_a^2}\;,\\
\mathcal{M}_{+-+-}^{\rm ALP}&=&-{g_{a\gamma\gamma}^2\over 4}{u^2\over u-m_a^2}\;.
\end{eqnarray}
Again, after combining them with the complete QED helicity amplitudes, we can get the total cross section valid in high-energy regime.

\begin{figure}[htb!]
\begin{center}
\includegraphics[width=0.48\textwidth]{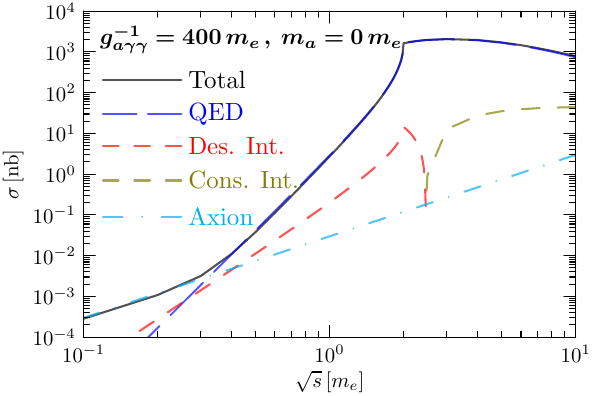}
\;
\includegraphics[width=0.48\textwidth]{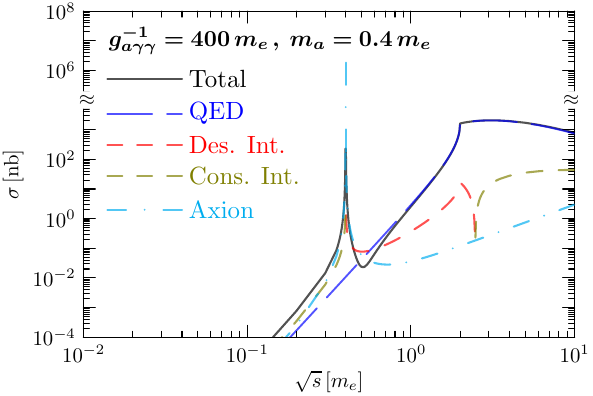}
\\[1em]
\includegraphics[width=0.48\textwidth]{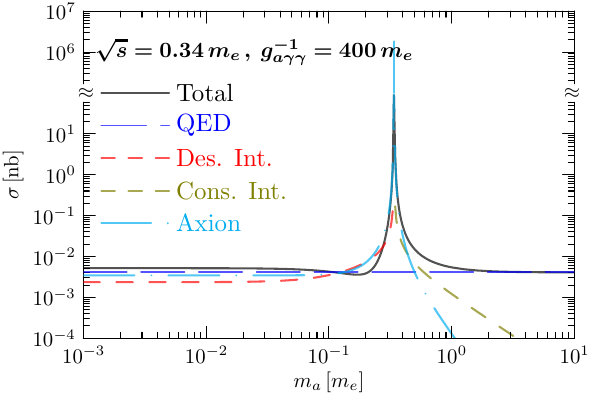}
\caption{The light-by-light scattering cross sections in ALP theory as a function of c.m. energy $\sqrt{s}$ (top panels) and $m_a$ (bottom panel). They include QED contribution (blue long-dashed line), ALP contribution (cyan dash-dotted line), destructive interference (red dashed line), constructive interference (olive dashed line) and total cross section (black solid line). For illustration, we fix $g_{a\gamma\gamma}^{-1}=400m_e$, $m_a=0$ and $g_{a\gamma\gamma}^{-1}=400m_e$, $m_a=0.4 m_e$ ($\sqrt{s}=0.34m_e$, $g_{a\gamma\gamma}^{-1}=400m_e$) in the up (bottom) panels.
}
\label{fig:xsALP}
\end{center}
\end{figure}
In Fig.~\ref{fig:xsALP}, we display the light-by-light scattering cross sections in ALP theory as a function of c.m. energy $\sqrt{s}$ (top panels) and $m_a$ (bottom panel).
We fix $g_{a\gamma\gamma}^{-1}=400m_e$, $m_a=0$ and $g_{a\gamma\gamma}^{-1}=400m_e$, $m_a=0.4 m_e$ ($\sqrt{s}=0.34m_e$, $g_{a\gamma\gamma}^{-1}=400m_e$) in the up (bottom) panels for illustration. The adopted values of $g_{a\gamma\gamma}^{-1}$ here and below are just benchmark points showing the ALP cross section together with the QED one in the same figure. We will explore the exclusion limit of $g_{a\gamma\gamma}$ in next section. Note that we take $m_a=0$ as an example of light ALP with $m_a\ll m_e$. We will scan the ALP mass and keep the calculated decay width $\Gamma_a$ in cross section when exploring the exclusion limit of $m_a$ and $g_{a\gamma\gamma}$ below. When $m_a$ crosses $\sqrt{s}$ from above to below, the interference term changes from constructive to destructive. The ALP contribution exhibits a resonance enhancement near $\sqrt{s}\approx m_a$, so does the total cross section.
On the other hand, if the ALP is produced on-shell with $\sqrt{s}=m_a$, it can become long-lived or decay to diphoton in the relevant parameter space. In case of on-shell production, we have $s=m_a^2$ and the ALP cross section is simply given as
\begin{equation}
\sigma^{\rm ALP}\equiv\sigma_{\gamma\gamma\to a}
=
{ \pi g_{a\gamma\gamma}^2 m_a^2\over 8 } \frac{1}{2E_a} \delta(E_a - m_a)
=
{ \pi g_{a\gamma\gamma}^2 m_a^2\over 16 E_a^2 }
=
{ \pi  \over 16  } g_{a\gamma\gamma}^2
\end{equation}
which only depends on the coupling constant $g_{a\gamma\gamma}$ and has no interference with pure QED contribution.

\begin{figure}[htb!]
\begin{center}
\includegraphics[width=0.48\textwidth]{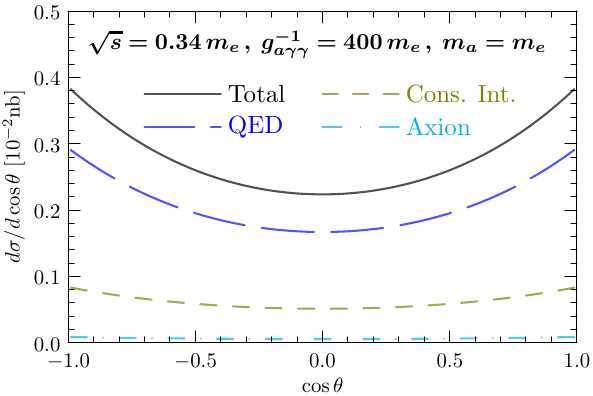}
\;
\includegraphics[width=0.48\textwidth]{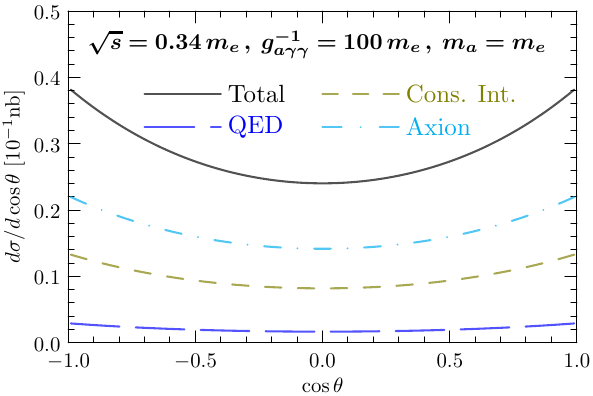}
\\
\includegraphics[width=0.48\textwidth]{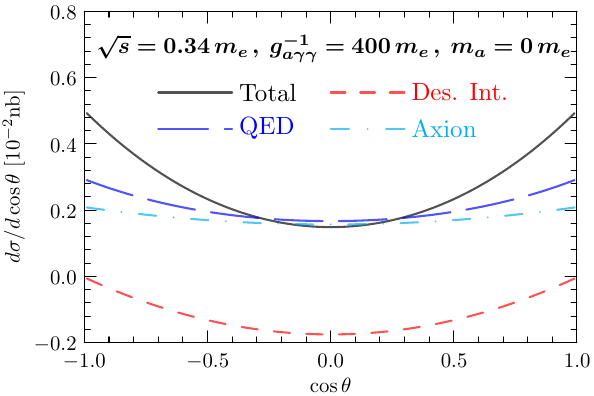}
\;
\includegraphics[width=0.48\textwidth]{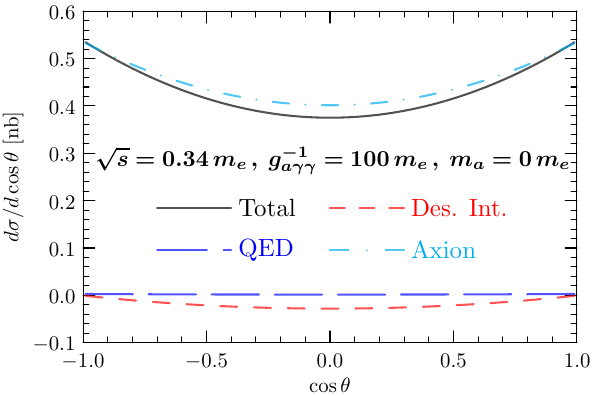}
\caption{The differential cross sections $d\sigma/d\cos\theta$ in ALP theory with $\sqrt{s}=0.34m_e$. For illustration, we choose $g_{a\gamma\gamma}^{-1}=400m_e$ and $m_a=m_e$ (top left panel), $g_{a\gamma\gamma}^{-1}=100m_e$ and $m_a=m_e$ (top right panel), $g_{a\gamma\gamma}^{-1}=400m_e$ and $m_a=0$ (bottom left panel), and $g_{a\gamma\gamma}^{-1}=100m_e$ and $m_a=0$ (bottom right panel). They include QED contribution (blue long-dashed line), ALP contribution (cyan dash-dotted line), constructive interference (olive dashed line), destructive interference (red dashed line), and total cross section (black solid line).
}
\label{fig:xcosALP}
\end{center}
\end{figure}
Fig.~\ref{fig:xcosALP} shows the differential cross section $d\sigma/d\cos\theta$ with $\sqrt{s}=0.34m_e$. 
For illustration, we choose $g_{a\gamma\gamma}^{-1}=400m_e$ and $m_a=m_e$ (top left panel), $g_{a\gamma\gamma}^{-1}=100m_e$ and $m_a=m_e$ (top right panel), $g_{a\gamma\gamma}^{-1}=400m_e$ and $m_a=0$ (bottom left panel), and $g_{a\gamma\gamma}^{-1}=100m_e$ and $m_a=0$ (bottom right panel). As expected, the ALP contribution increases with growing coupling $g_{a\gamma\gamma}$. Pure EH (ALP) contribution dominates the total cross section for small (large) $g_{a\gamma\gamma}$. Changing $m_a$ from $m_e$ to zero shifts the interference term from constructive to destructive.

\begin{figure}[ht]
\begin{center}
\includegraphics[width=0.48\textwidth]{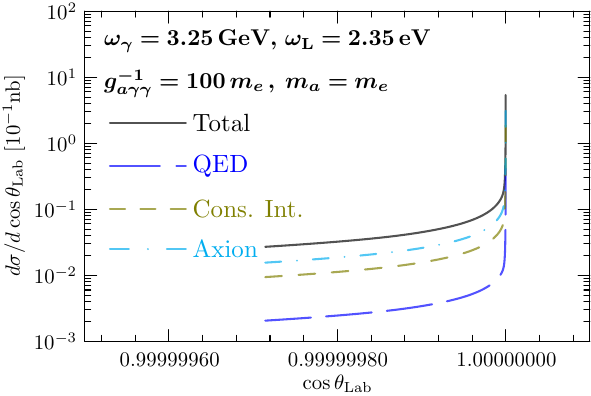}
\;
\includegraphics[width=0.48\textwidth]{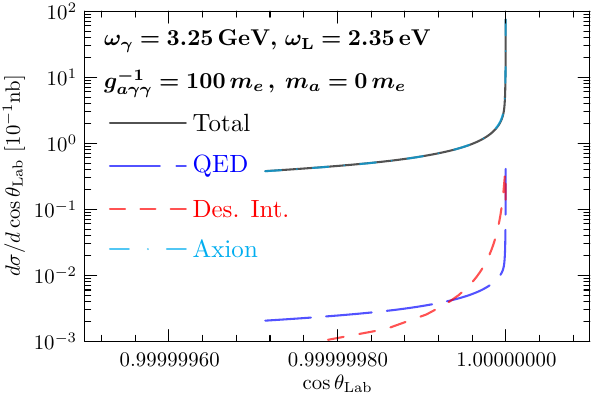}
\caption{
The differential cross sections $d\sigma/d\cos\theta_{\rm Lab}$ in ALP theory with $\omega_\gamma =3.25 {\rm GeV}$, $\omega_L =2.35 {\rm eV}$ and $\vartheta=10^\circ$ with coupling $g_{a\gamma\gamma}^{-1}=100m_e$ and mass parameters $m_a=m_e$ (left) and $m_a=0$ (right). They include QED contribution (blue long-dashed line), ALP contribution (cyan dash-dotted line), constructive interference (olive dashed line), destructive interference (red dashed line), and the total contribution (black solid line).
}
\label{fig:xcoslab:ALP:xs}
\end{center}
\end{figure}
Fig.~\ref{fig:xcoslab:ALP:xs} shows the differential cross sections $d\sigma/d\cos\theta_{\rm Lab}$ with incoming photon energy $\omega_\gamma =3.25 \,{\rm GeV}$, $\omega_L =2.35 \,{\rm eV}$ (green laser) and an intersection angle $\vartheta=10^\circ$,
which corresponds to
a c.m. energy $\sqrt{s} \approx 0.34m_e$.
The left and right panels show the cases with coupling $g_{a\gamma\gamma}^{-1}=100m_e$ and mass parameters $m_a=m_e$ and $m_a=0$, respectively.
Similar to the case of BI theory,
most of the outgoing photons are collinear with the incoming gamma-ray, and there exists difference between the QED background and the ALP correction in the forward region.
For the case $m_a=0$, the differences are relatively larger compared to the BI theory. However,
an order of $10^{-4} \sim 10^{-3}$ uncertainty for the polar angle measurement is still necessary in order to observe it in the experiment.

\section{Sensitivity of laser-assisted light-by-light scattering to BI and ALP parameters}
\label{sec:Sen}

We use the following significance formula to estimate the sensitivity of laser-assisted light-by-light scattering to BI and ALP parameters
\begin{eqnarray}
\label{eq:sens}
\mathcal{S}={|\sigma^{\rm Total}-\sigma^{\rm QED}|\cdot \mathcal{L}\over \sqrt{\sigma^{\rm Total}\cdot \mathcal{L}}} \sim {S\over \sqrt{S+B}}\;,
\end{eqnarray}
where $\mathcal{S}$ denotes the significance, $\mathcal{L}$ is the integrated luminosity, $\sigma^{\rm Total}=\sigma^{\rm QED+BI}$ ($\sigma^{\rm QED+ALP}$) in BI (ALP) theory, and $S~(B)$ is the number of signal (background) events.
In the above calculation of the significance $\mathcal{S}$, systematic
uncertainties have been neglected. We consider only the statistical uncertainties given by $\sqrt{S+B}$.
The exclusion limit is obtained by requiring a $2\sigma$ significance, {\it i.e.} $\mathcal{S}=2$, which corresponds an about 95\% CL exclusion limit.
Furthermore, in the calculation of the significance $\mathcal{S}$, only the irreducible contribution from the SM light-by-light scattering is considered as the background. In principle there can be other backgrounds from reducible processes. For instance, two of the photons in the multiple photon production --- such as $\gamma\gamma\to\gamma\gamma\gamma\gamma$, are not recorded by the detector.
When the c.m. energy exceeds the electron-positron pair production threshold, the process $\gamma\gamma\to e^{+}e^{-}$ can also contribute if the electron-positron pair is misidentified as two photons. Additionally,
the process $\gamma\gamma\to e^{+}e^{-}\gamma\gamma$ can also be misidentified as two-photon production when the electron-positron pair is not recorded by the detector.
However, the contributions of all these processes are reducible and cannot dominate over the irreducible contribution. Considering only the irreducible background is sufficient for a primary estimation of the experimental sensitivity.

The luminosity is given by~\cite{Herr:2003em,Cannoni:2016hro}
\begin{eqnarray}
\mathcal{L}=N_\gamma \rho_\omega \ell N_b f t\;,
\end{eqnarray}
where $N_\gamma\approx 2.3\times 10^8$ denotes the number of energetic photons in a bunch passing through the collimator~\cite{Sangal:2021qeg}, $\rho_\omega={m_e^2 \eta^2 \omega_L\over 4\pi e^2}$ with $\eta$ (taken as 1) being the intensity parameter is the laser photon density~\cite{Greiner:1992bv}, $N_b = 2700$ is the number of individual bunches in the electron beam of the European XFEL accelerator used for the LUXE experiment~\cite{LUXE:2023crk}, $f = 1~{\rm Hz}$ is the laser operating frequency~\cite{LUXE:2023crk}, $\ell=9~{\rm \mu m}$ is the gamma photon pathlength through the laser focus (corresponding to 30 fs duration~\cite{Yakimenko:2019sya}), and $t=5\times 10^6~{\rm s}$ is the physics data-taking time~\cite{LUXE:2023crk}.
It corresponds to an integrated luminosity of $\mathcal{L}\approx 0.2~{\rm ab}^{-1}$. The main high-power laser facilities in operation or in design~\cite{Sarri:2025qng} provide much larger intensity parameter $\eta$ and thus we take a greater integrated luminosity as $\mathcal{L}= 1~{\rm ab}^{-1}$ in the significance.

\begin{figure}[htb!]
\begin{center}
\includegraphics[width=0.48\textwidth]{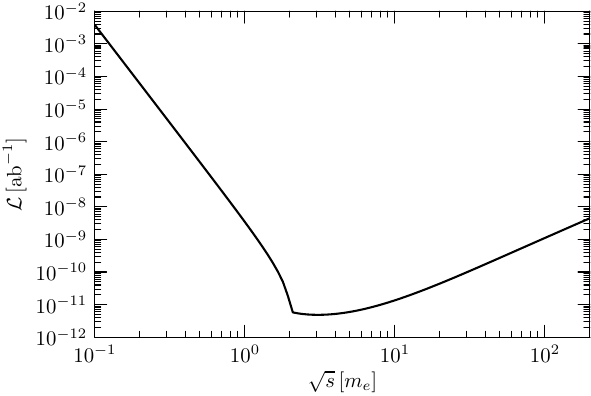}
\hfill
\includegraphics[width=0.48\textwidth]{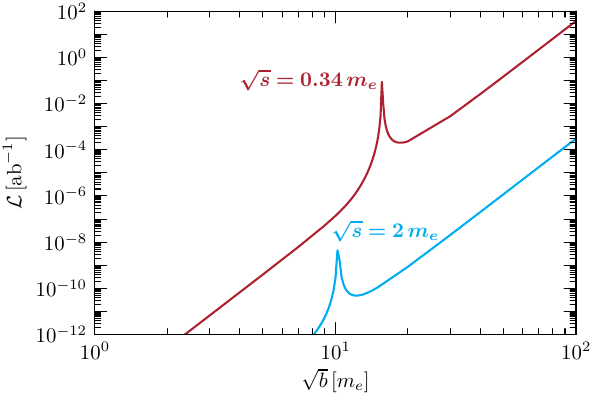}
\caption{The required luminosity to observe 10 events of SM light-by-light scattering (left) and to reach 95\% CL exclusion for laser-assisted light-by-light scattering as a function of $\sqrt{b}$ in BI theory (right). In the right panel, we choose $\sqrt{s}=0.34m_e$ (dark red line) and $2m_e$ (cyan line) for illustration.
}
\label{fig:excl:bi}
\end{center}
\end{figure}

For completeness, we first estimate the observation potential of laser-assisted light-by-light scattering in the SM by requiring the number of events to be 10.
The left panel of Fig.~\ref{fig:excl:bi} shows the contour in the plane of the c.m. energy and the required luminosity. One can see that below the electron-positron pair production threshold,
the observation potential increases dramatically with increasing $\sqrt{s}$. Then, it drops relatively slowly above the threshold. In the case of $\sqrt{s}=m_e$, only $\sim 1\,{\rm nb}^{-1}$ luminosity is needed to observe the SM light-by-light processes.

The right panel of Fig.~\ref{fig:excl:bi} shows the required luminosity to reach 95\% exclusion limit as a function of $\sqrt{b}$. We choose $\sqrt{s}=0.34m_e$ (dark red) and $2m_e$ (cyan) for illustration.
It turns out that as $\sqrt{b}$ increases or $\sqrt{s}$ decreases, the required luminosity rises up to 10 ${\rm ab}^{-1}$ for $\sqrt{s}=0.34m_e$ and $\sqrt{b}=100m_e$. At some critical $\sqrt{b}$ values, e.g., $\sqrt{b}\approx 15m_e$ for $\sqrt{s}=0.34 m_e$ or $\sqrt{b}\approx 10m_e$ for $\sqrt{s}=2m_e$, the total cross section reaches its minimum as displayed in the right panel of Fig.~\ref{fig:xsBI} and one requires a peaked luminosity to obtain the same exclusion significance.

We then compare the above exclusion from laser-assisted light-by-light scattering with the existing limits on the BI $\sqrt{b}$ parameter from the LHC~\cite{Ellis:2017edi}. First of all, the diphoton system at the LHC is required to have a much higher invariant mass $m_{\gamma\gamma}$ ($>6$ GeV or $>25$ GeV) than our case, at which the cross section in BI theory dominates over the QED cross section. The fiducial cross section for light-by-light scattering in relativistic heavy-ion collisions is about 65 nb for $m_{\gamma\gamma}>6$ GeV and 2 nb for $m_{\gamma\gamma}>25$ GeV. They translate into the limit $\sqrt{b}\gtrsim 100$ GeV and 210 GeV, respectively.
These bounds agree with the implicit condition $\sqrt{b} \gg m_{\gamma\gamma}$ in order to have a safe perturbative expansion of the nonlinear BI Lagrangian.
The sensitivity region of $\sqrt{b}$ at LHC is much larger than the range of interest in Fig.~\ref{fig:excl:bi}. Thus, the laser-assisted light-by-light scattering with low $\sqrt{s}$ is sensitive to small $\sqrt{b}$ parameter region and provides a complementary constraint on BI theory for low invariant mass $m_{\gamma\gamma}$.

\begin{figure}[htb!]
\begin{center}
\includegraphics[width=0.68\textwidth]{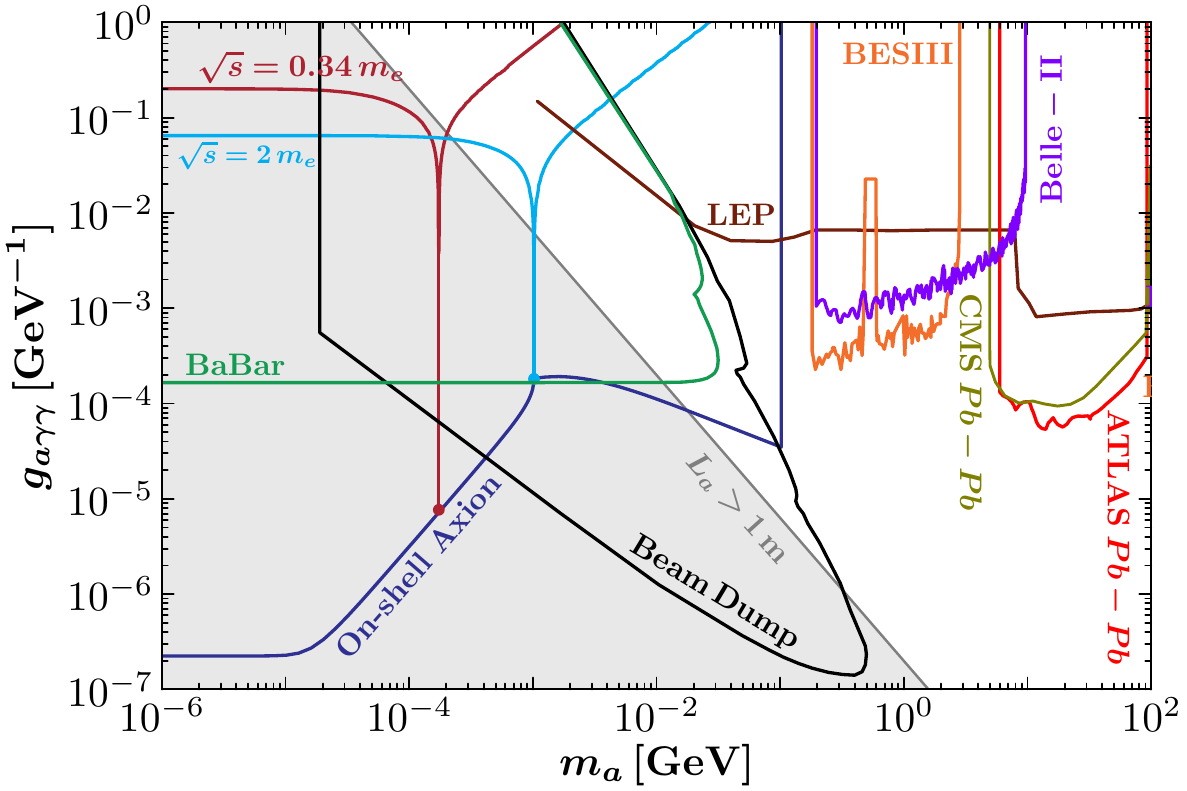}
\caption{Sensitivity of laser-assisted light-by-light scattering to ALP-photon coupling $g_{a\gamma\gamma}$ with $\sqrt{s}=0.34m_e$ (dark red line) and $2m_e$ (cyan line). The sensitivity from on-shell ALP production is shown in dark blue line. For comparison, we also show the exclusion limits from LEP~\cite{Jaeckel:2015jla} (brown), BESIII~\cite{BESIII:2022rzz,BESIII:2024hdv} (orange), Belle II~\cite{Belle-II:2020jti} (purple), ATLAS PbPb~\cite{ATLAS:2020hii} (red), CMS PbPb~\cite{CMS:2024bnt} (olive), beam dump (black, combined limit from CHARM~\cite{BERGSMA1985458}, E141~\cite{PhysRevLett.59.755}, E137~\cite{Dolan:2017osp,PhysRevD.38.3375}, NuCal~\cite{Dobrich:2019dxc,Blumlein:1990ay}, NA64~\cite{NA64:2020qwq}. see also the summaries~\cite{Dobrich:2019dxc,dEnterria:2021ljz}), and BaBar~\cite{BaBar:2017tiz,Dolan:2017osp} (green). The decay length of ALP $L_a$ is greater than 1 meter in the gray region.}
\label{fig:excl:ax}
\end{center}
\end{figure}
In Fig.~\ref{fig:excl:ax}, we also show the sensitivity of laser-assisted light-by-light scattering to ALP-photon coupling $g_{a\gamma\gamma}$ with $\sqrt{s}=0.34m_e$ (dark red line) and $2m_e$ (cyan line). One can see that, except at the resonance points $\sqrt{s}=m_a$, the ALP is produced off-shell and the reachable limits are complementary to those from other high-energy colliders -- such as LEP, BESIII, Belle II, and the LHC -- in the low-mass region. However, these sensitivity regions are excluded by the BaBar search for invisible ALP which is long-lived enough to escape detection~\cite{BaBar:2017tiz,Dolan:2017osp}.

On the other hand, the on-shell production of $a$,
$\gamma\gamma \to a$, has very large cross section. The produced ALP can be long-lived to escape detection or decay into diphoton with 100\% branching fraction. In either case, we have $\sigma^{\rm Total}=\sigma^{\rm QED}+\sigma^{\rm ALP}$ in the above significance formula because there is no interference between ALP production and QED scattering.
However, in the parameter region of small $m_a$, the ALP is nearly invisible, and hence it is challenging in practice to identify such events. There are two practical approaches to observe the on-shell production of an invisible ALP: (1) using polarized initial photons;
(2) measuring energy loss of the total system or either one of the two initial photon beams.
For the first case, because only the photons with the same helicity states (in the c.m. frame) can annihilate (for QED background both helicity configurations can have non-trivial scattering), the polarizations of outgoing photons and remaining beams receive non-trivial corrections. However, this approach is challenged by the difficulty of measuring the polarization of the gamma-ray (due to its high energy) and that of the laser (due to its high intensity).
On the other hand, the second approach is based on energy conservation. The energy loss, which is defined as $E_{\rm Loss}^{\rm Exp} = E_{\rm in}^{\rm Exp} - E_{\rm rem}^{\rm Exp}$,
where $E_{\rm in}^{\rm Exp}$ is the injection energy of the two incoming photon beams (or either one of the beams) and $E_{\rm rem}^{\rm Exp}$ is the remaining energy of two incoming photon beams (or either one of the beams)~\footnote{In practice, beam remnants after collision can either be collected for further use or simply dumped. Particularly, laser beam remnants can be easily recollimated by an off-axis paraboloid (OAP) mirror to return to the laser room for subsequent use~\cite{Bamber:1999zt}. This provides a straightforward method to measure the energy imbalance by analyzing only the laser beam.}, should be determined by the pure QED prediction $E_{\rm QED}^{\rm The}$, {\it i.e.}, $E_{\rm Loss}^{\rm Exp} - E_{\rm QED}^{\rm The} \approx 0$.
If there is on-shell ALP production, we then have $E_{\rm Loss}^{\rm Exp} - E_{\rm QED}^{\rm The} \approx E_{\rm ALP}^{\rm The}$,
where $E_{\rm ALP}^{\rm The}$ stands for theoretical prediction of the energy carried away by the invisible ALP.
Assuming that the injection energy $E_{\rm in}^{\rm Exp}$ is a constant, the uncertainty of energy loss measurement is determined by $E_{\rm QED}^{\rm The}$ (more precisely $E_{\rm QED}^{\rm Exp}$), and hence is proportional to the square root of pure QED event number $\sqrt{N_{\rm QED}^{\rm Exp}}$. In this sense, our sensitivity estimation formula in
Eq.~(\ref{eq:sens}) is still valid.

The sensitivity from on-shell ALP production is shown by dark blue line in Fig.~\ref{fig:excl:ax}. In this case we take $\sqrt{s}=m_a$ less than muon lepton mass. Due to the sizable on-shell production cross section, one can see that the reachable limit of $g_{a\gamma\gamma}\sim 2\times 10^{-7}~{\rm GeV}^{-1}$ is three orders of magnitude smaller than BaBar exclusion limit for $m_a\lsim 10^{-5}~{\rm GeV}$. In high mass range of $10^{-2}-10^{-1}~{\rm GeV}$, compared to other collider and beam dump experiments, we find that the on-shell ALP production provides complementary search potential.

\section{Conclusion}
\label{sec:Con}

The precision measurements of well-known light-by-light reactions lead to important insights of QED vacuum polarization. The laser of an
intense electromagnetic field strength provides an essential tool for exploring strong-field QED and new physics beyond SM in the high-precision frontier. In this work, we investigate the low-energy light-by-light scattering in the collision of a photon beam and a laser pulse of high-intensity. We propose to search for the impact of BI and ALP theories in laser-assisted light-by-light scattering.

We take into account the QED light-by-light scattering $\gamma\gamma\to \gamma\gamma$ with an incoming photon being a classical background field. The scattering cross sections are calculated using complete QED helicity amplitudes. We then combine them with the amplitudes in BI or ALP theory to evaluate the total cross section. The laser-assisted SM light-by-light scattering should be observable in future experiments with very moderate integrated luminosities. The sensitivity of laser-assisted light-by-light scattering to BI and ALP parameters is presented. We find that
\begin{itemize}
\item The total cross section of QED+BI for $\sqrt{s}<2m_e$ decreases as $\sqrt{b}$ increases and reaches a minimum value at some critical point because of the negative interference term. For $\sqrt{b}<10^2 m_e$, as $\sqrt{b}$ increases or $\sqrt{s}$ decreases, the required luminosity to reach 95\% CL exclusion limit arises up to $10~{\rm ab}^{-1}$ and needs a peaked value to obtain the same exclusion significance at the critical $\sqrt{b}$.
\item The ALP contribution
exhibits a resonance enhancement for the light-by-light cross section near $\sqrt{s}\approx m_a$. Except at the resonance points $\sqrt{s}=m_a$, the ALP is produced off-shell and the reachable limits are complementary to those from other high-energy colliders in the low-mass region. However, these sensitivity regions are excluded by the BaBar search for invisible ALP.
\item The on-shell ALP production with $\sqrt{s}=m_a$ has sizable production cross section. The reachable limit of $g_{a\gamma\gamma}\sim 2\times 10^{-7}~{\rm GeV}^{-1}$ for $m_a\lsim 10^{-5}~{\rm GeV}$ is three orders of magnitude smaller than BaBar exclusion limit. In high mass range of $10^{-2}-10^{-1}~{\rm GeV}$, the on-shell ALP production provides complementary search potential to beam dump experiments.
\end{itemize}

\acknowledgments

We would like to thank Liangliang Ji and Zhan Bai for useful discussions.
T.~L. is supported by the National Natural Science Foundation of China (Grant No. 12375096, 12035008, 11975129). K. M. is supported by the Natural Science Basic Research Program of Shaanxi (Program No. 2023-JC-YB-041).

\appendix

\section{Lorentz transformation to the laboratory (Lab) frame}
\label{app:Lab}

The laboratory frame is defined as
\bea
p_\gamma &=& \omega_\gamma(1,\, 0,\, 0,\, 1) \,,
\\
p_L &=& \omega_L(1,\, \sin\vartheta,\, 0,\, -\cos\vartheta) \,.
\ena
Then, the total momentum is $P = p_\gamma + p_L = \sqrt{s} $, and its invariant mass is given as
\begin{equation}
P^2 = s = 2 p_\gamma \cdot p_L = 2  \omega_\gamma \omega_L (1 + \cos\vartheta)
= 4  \omega_\gamma \omega_L \cos^2\frac{\vartheta}{2}\;.
\end{equation}
The total momentum in the Lab frame can be written as
\begin{equation}
P = E_s (1,\, \beta_s\sin\vartheta_s,\, 0,\, \beta_s\cos\vartheta_s ) \,,
\end{equation}
where
\bea
E_s &=& \omega_\gamma + \omega_L\;,
\\
\gamma_s &=& \frac{ \omega_\gamma + \omega_L }{ 2 \cos\dfrac{\vartheta}{2} \sqrt{\omega_\gamma \omega_L} }\;,
\\
\beta_s^2 &=& 1 - \frac{1 }{ \gamma_s^2 }
=
1 -  \frac{ 4 \omega_\gamma \omega_L \cos^2\dfrac{\vartheta}{2} }{ (\omega_\gamma + \omega_L)^2 }
=
 \frac{ \omega_\gamma^2 - 2\omega_\gamma \omega_L \cos\vartheta + \omega_L^2 }{ (\omega_\gamma + \omega_L)^2 }\;,
\\
\cos\vartheta_s &=& \frac{ \omega_\gamma - \omega_L\cos\vartheta  }{ \beta_s E_s  }\;.
\ena
The polar angle can be simplified as
\begin{equation}
\cos\vartheta_s
=
\frac{ (\omega_\gamma - \omega_L \cos\vartheta) }{ (\omega_\gamma^2 - 2\omega_\gamma \omega_L \cos\vartheta + \omega_L^2)^{1/2} }\;.
\end{equation}

Now we can move to the rest frame of $P$
\begin{equation}
P^\ast = \sqrt{s}(1,\, 0,\, 0,\, 0)\;.
\end{equation}
This can be achieved by a rotation $R_Y(-\vartheta_s)$ and a boost $R_Z(-\beta_s)$
\begin{equation}
R_Y(-\vartheta_s)
=
\left(\begin{array}{cccc}
1 & 0 & 0 & 0
\\
0 & \cos\vartheta_s  & 0 & -\sin\vartheta_s
\\
0 & 0 & 1 & 0
\\
0 & \sin\vartheta_s  & 0 & \cos\vartheta_s
\end{array}\right)
\;,\quad
R_Z(-\beta_s)
=
\left(\begin{array}{cccc}
\gamma_s & 0 & 0 & -\gamma_s\beta_s
\\
0 & 1 & 0 & 0
\\
0 & 0 & 1 & 0
\\
-\gamma_s\beta_s & 0  & 0 & \gamma_s
\end{array}\right)\;.
\end{equation}
In this reference frame we have
\bea
p_\gamma^\ast &=& R_Z(-\beta_s) R_Y(-\vartheta_s) p_\gamma
\nonumber\\
&=&
R_Z(-\beta_s) \omega_\gamma(1,\, -\sin\vartheta_s,\, 0,\, \cos\vartheta_s)
\nonumber\\
&=&
\gamma_s \omega_\gamma(1 - \beta_s \cos\vartheta_s ,\, -\sin\vartheta_s,\, 0,\,  \cos\vartheta_s -\beta_s )
\ena
and
\bea
p_L^\ast &=& R_Z(-\beta_s) R_Y(-\vartheta_s) p_L
\nonumber
\\
&=&
R_Z(-\beta_s) \omega_L(1,\, \sin(\vartheta+\vartheta_s),\, 0,\, -\cos(\vartheta+\vartheta_s))
\nonumber
\\
&=&
\gamma_s \omega_L(1 + \beta_s \cos(\vartheta+\vartheta_s),\, \sin(\vartheta+\vartheta_s),\, 0,\, -\beta_s - \cos(\vartheta+\vartheta_s) )
\ena

Now, we can obtain the energy and polar angle of the gamma-ray as
\bea
\omega_\gamma^\ast &=& \gamma_s \omega_\gamma(1 - \beta_s \cos\vartheta_s )\;,
\\[3mm]
\cos\vartheta_\gamma^\ast &=& \frac{ \cos\vartheta_s -\beta_s }{ 1 - \beta_s \cos\vartheta_s  }\;,
\\[3mm]
\sin\vartheta_\gamma^\ast &=& \frac{ -\sin\vartheta_s }{ 1 - \beta_s \cos\vartheta_s  }\;.
\ena
The momentum $p^\ast_\gamma$ is parameterized as
\begin{equation}
p^\ast_\gamma = \omega_\gamma^\ast(1,\, \sin\vartheta_\gamma^\ast ,\, 0,\, \cos\vartheta_\gamma^\ast ) \,.
\end{equation}

A further rotation $ R_Y(-\vartheta_\gamma^\ast) $ will leave the momentum $p^\ast_\gamma $ and $p^\ast_L $ along the $z$-axis,
and $p^\ast_\gamma $ is along the positive axis.
We denote this reference frame as $R^\star$,
in which the scattering amplitudes are calculated.
Because of the azimuthal angle invariance in this reference frame,
only polar angle ($\theta^\star$) of the outgoing photon is non-trivial,
and the momentum is parameterized as
\begin{equation}
q^\star = \omega^\star(1,\, \sin\theta^\star ,\, 0,\, \cos\theta^\star ) \,.
\end{equation}
It is related to the Lab frame momentum as
\begin{equation}
q^\star = R_Y(-\vartheta_\gamma^\ast) R_Z(-\beta_s) R_Y(-\vartheta_s)  q^\star_{Lab}
\end{equation}
or inversely,
\bea
q^\star_{Lab}
&=&
R_Y(\vartheta_s) R_Z(\beta_s) R_Y(\vartheta_\gamma^\ast) q^\star
\nonumber\\[3mm]
&=&
R_Y(\vartheta_s) R_Z(\beta_s) \omega^\star(1,\, \sin(\theta^\star+\vartheta_\gamma^\ast),\, 0,\, \cos(\theta^\star+\vartheta_\gamma^\ast) )
\nonumber\\[3mm]
&=&
R_Y(\vartheta_s) \gamma_s \omega^\star
(1+ \beta_s \cos(\theta^\star+\vartheta_\gamma^\ast),\,
\sin(\theta^\star+\vartheta_\gamma^\ast),\,
0,\,
\cos(\theta^\star+\vartheta_\gamma^\ast) + \beta_s
)\;.
\ena
After the final rotation, we have
\bea
q^\star_{Lab, 0}
&=&
\gamma_s \omega^\star \big[1+ \beta_s \cos(\theta^\star+\vartheta_\gamma^\ast) \big]\;,
\\[3mm]
q^\star_{Lab, 1}
&=&
\gamma_s \omega^\star \big[
\cos\vartheta_s\sin(\theta^\star+\vartheta_\gamma^\ast)
+
\sin\vartheta_s ( \cos(\theta^\star+\vartheta_\gamma^\ast) + \beta_s )
\big]\;,
\\[3mm]
q^\star_{Lab, 2}
&=&
0\;,
\\[3mm]
q^\star_{Lab, 3}
&=&
\gamma_s \omega^\star \big[
\sin\vartheta_s\sin(\theta^\star+\vartheta_\gamma^\ast)
+
\cos\vartheta_s ( \cos(\theta^\star+\vartheta_\gamma^\ast) + \beta_s )
\big]\;.
\ena
Then we can obtain the polar angle of outgoing photon in the Lab frame as
\begin{equation}
\cos\theta_{Lab}
=
\frac{ \sin\vartheta_s\sin(\theta^\star+\vartheta_\gamma^\ast)
+
\cos\vartheta_s ( \cos(\theta^\star+\vartheta_\gamma^\ast) + \beta_s ) }{ 1+ \beta_s \cos(\theta^\star+\vartheta_\gamma^\ast) }
\end{equation}
which can be further simplified as
\begin{equation}
\cos\theta_{Lab}
=
\frac{ \cos(\theta^\star+\vartheta_\gamma^\ast - \vartheta_s) + \beta_s \cos\vartheta_s }
{ 1+ \beta_s \cos(\theta^\star+\vartheta_\gamma^\ast) }\;.
\end{equation}
In case of $\vartheta_s = 0$, we have
\begin{equation}
\cos\theta_{Lab}
=
\frac{ \cos(\theta^\star+\vartheta_\gamma^\ast) + \beta_s  }
{ 1+ \beta_s \cos(\theta^\star+\vartheta_\gamma^\ast) }\;.
\end{equation}
One can see that it is consistent with a single boost.
In case of $\beta_s = 0$, we have
\begin{equation}
\cos\theta_{Lab}
=
\cos(\theta^\star+\vartheta_\gamma^\ast)\;.
\end{equation}
It turns out that $\theta^\star$ is just the corresponding polar angle, i.e.,
the c.m. frame and the Lab frame are the same (the angle $\vartheta_\gamma^\ast$
is trivial and can be removed by the redefinition of $z$-axis).
This is a roughly check of above calculations.

In the limit $\beta_s \to 1$, we have
\begin{equation}
\cos\theta_{Lab}
\to
\frac{ \cos(\theta^\star+\vartheta_\gamma^\ast - \vartheta_s) + \cos\vartheta_s }
{ 1+ \cos(\theta^\star+\vartheta_\gamma^\ast) }
\to 1 \quad\text{for}\quad \vartheta_s \to 0\;.
\end{equation}
The above limit does not depend on $\theta^\star$.

\bibliographystyle{JHEP}
\bibliography{refs}

\providecommand{\href}[2]{#2}\begingroup\raggedright\begin{thebibliography}{100}

\bibitem{Heisenberg:1936nmg}
W.~Heisenberg and H.~Euler, \emph{{Consequences of Dirac's theory of
  positrons}}, \href{http://dx.doi.org/10.1007/BF01343663}{\emph{Z. Phys.} {\bf
  98} (1936) 714--732}, [\href{http://arxiv.org/abs/physics/0605038}{{\tt
  physics/0605038}}].

\bibitem{Weisskopf:1936hya}
V.~Weisskopf, \emph{{The electrodynamics of the vacuum based on the quantum
  theory of the electron}}, {\emph{Kong. Dan. Vid. Sel. Mat. Fys. Med.} {\bf
  14N6} (1936) 1--39}.

\bibitem{Dunne:2004nc}
G.~V. Dunne, \emph{{Heisenberg-Euler effective Lagrangians: Basics and
  extensions}}, pp.~445--522.
\newblock 6, 2004.
\newblock \href{http://arxiv.org/abs/hep-th/0406216}{{\tt hep-th/0406216}}.
\newblock 10.1142/9789812775344\_0014.

\bibitem{Karplus:1950zz}
R.~Karplus and M.~Neuman, \emph{{The scattering of light by light}},
  \href{http://dx.doi.org/10.1103/PhysRev.83.776}{\emph{Phys. Rev.} {\bf 83}
  (1951) 776--784}.

\bibitem{dEnterria:2013zqi}
D.~d'Enterria and G.~G. da~Silveira, \emph{{Observing light-by-light scattering
  at the Large Hadron Collider}},
  \href{http://dx.doi.org/10.1103/PhysRevLett.111.080405}{\emph{Phys. Rev.
  Lett.} {\bf 111} (2013) 080405}, [\href{http://arxiv.org/abs/1305.7142}{{\tt
  1305.7142}}].

\bibitem{ATLAS:2017fur}
{\scshape ATLAS} collaboration, M.~Aaboud et~al., \emph{{Evidence for
  light-by-light scattering in heavy-ion collisions with the ATLAS detector at
  the LHC}}, \href{http://dx.doi.org/10.1038/nphys4208}{\emph{Nature Phys.}
  {\bf 13} (2017) 852--858}, [\href{http://arxiv.org/abs/1702.01625}{{\tt
  1702.01625}}].

\bibitem{CMS:2018erd}
{\scshape CMS} collaboration, A.~M. Sirunyan et~al., \emph{{Evidence for
  light-by-light scattering and searches for axion-like particles in
  ultraperipheral PbPb collisions at $\sqrt{s_\mathrm{NN}} =$ 5.02 TeV}},
  \href{http://dx.doi.org/10.1016/j.physletb.2019.134826}{\emph{Phys. Lett. B}
  {\bf 797} (2019) 134826}, [\href{http://arxiv.org/abs/1810.04602}{{\tt
  1810.04602}}].

\bibitem{ATLAS:2020hii}
{\scshape ATLAS} collaboration, G.~Aad et~al., \emph{{Measurement of
  light-by-light scattering and search for axion-like particles with 2.2
  nb$^{-1}$ of Pb+Pb data with the ATLAS detector}},
  \href{http://dx.doi.org/10.1007/JHEP03(2021)243}{\emph{JHEP} {\bf 03} (2021)
  243}, [\href{http://arxiv.org/abs/2008.05355}{{\tt 2008.05355}}].

\bibitem{CMS:2024bnt}
{\scshape CMS} collaboration, A.~Hayrapetyan et~al., \emph{{Measurement of
  light-by-light scattering and the Breit-Wheeler process, and search for
  axion-like particles in ultraperipheral PbPb collisions at
  $\sqrt{{s}_{\text{NN}}}$ = 5.02 TeV}},
  \href{http://dx.doi.org/10.1007/JHEP08(2025)006}{\emph{JHEP} {\bf 08} (2025)
  006}, [\href{http://arxiv.org/abs/2412.15413}{{\tt 2412.15413}}].

\bibitem{Wilson:1953zz}
R.~R. Wilson, \emph{{Scattering of 1.33 Mev Gamma-Rays by an Electric Field}},
  \href{http://dx.doi.org/10.1103/PhysRev.90.720}{\emph{Phys. Rev.} {\bf 90}
  (1953) 720--721}.

\bibitem{Jarlskog:1973aui}
G.~Jarlskog, L.~Joensson, S.~Pruenster, H.~D. Schulz, H.~J. Willutzki and G.~G.
  Winter, \emph{{Measurement of delbrueck scattering and observation of photon
  splitting at high energies}},
  \href{http://dx.doi.org/10.1103/PhysRevD.8.3813}{\emph{Phys. Rev. D} {\bf 8}
  (1973) 3813--3823}.

\bibitem{Schumacher:1975kv}
M.~Schumacher, I.~Borchert, F.~Smend and P.~Rullhusen, \emph{{Delbruck
  Scattering of 2.75-MeV Photons by Lead}},
  \href{http://dx.doi.org/10.1016/0370-2693(75)90685-1}{\emph{Phys. Lett. B}
  {\bf 59} (1975) 134--136}.

\bibitem{Akhmadaliev:1998zz}
S.~Z. Akhmadaliev et~al., \emph{{Delbruck scattering at energies of 140-450
  MeV}}, \href{http://dx.doi.org/10.1103/PhysRevC.58.2844}{\emph{Phys. Rev. C}
  {\bf 58} (1998) 2844--2850}.

\bibitem{Toll:1952rq}
J.~S. Toll, \emph{{The Dispersion relation for light and its application to
  problems involving electron pairs}},  other thesis, 1952.

\bibitem{Mignani:2016fwz}
R.~P. Mignani, V.~Testa, D.~G. Caniulef, R.~Taverna, R.~Turolla, S.~Zane
  et~al., \emph{{Evidence for vacuum birefringence from the first
  optical-polarimetry measurement of the isolated neutron star RX
  J1856.5\ensuremath{-}3754}},
  \href{http://dx.doi.org/10.1093/mnras/stw2798}{\emph{Mon. Not. Roy. Astron.
  Soc.} {\bf 465} (2017) 492--500},
  [\href{http://arxiv.org/abs/1610.08323}{{\tt 1610.08323}}].

\bibitem{Akhmadaliev:2001ik}
S.~Z. Akhmadaliev et~al., \emph{{Experimental investigation of high-energy
  photon splitting in atomic fields}},
  \href{http://dx.doi.org/10.1103/PhysRevLett.89.061802}{\emph{Phys. Rev.
  Lett.} {\bf 89} (2002) 061802},
  [\href{http://arxiv.org/abs/hep-ex/0111084}{{\tt hep-ex/0111084}}].

\bibitem{Sangal:2021qeg}
M.~Sangal, C.~H. Keitel and M.~Tamburini, \emph{{Observing light-by-light
  scattering in vacuum with an asymmetric photon collider}},
  \href{http://dx.doi.org/10.1103/PhysRevD.104.L111101}{\emph{Phys. Rev. D}
  {\bf 104} (2021) L111101}, [\href{http://arxiv.org/abs/2101.02671}{{\tt
  2101.02671}}].

\bibitem{Schwinger:1951nm}
J.~S. Schwinger, \emph{{On gauge invariance and vacuum polarization}},
  \href{http://dx.doi.org/10.1103/PhysRev.82.664}{\emph{Phys. Rev.} {\bf 82}
  (1951) 664--679}.

\bibitem{Greiner:1992bv}
W.~Greiner and J.~Reinhardt, \emph{{Quantum electrodynamics}}.
\newblock 1992.

\bibitem{Hartin:2018egj}
A.~Hartin, \emph{{Strong field QED in lepton colliders and electron/laser
  interactions}}, \href{http://dx.doi.org/10.1142/S0217751X18300119}{\emph{Int.
  J. Mod. Phys. A} {\bf 33} (2018) 1830011},
  [\href{http://arxiv.org/abs/1804.02934}{{\tt 1804.02934}}].

\bibitem{Fedotov:2022ely}
A.~Fedotov, A.~Ilderton, F.~Karbstein, B.~King, D.~Seipt, H.~Taya et~al.,
  \emph{{Advances in QED with intense background fields}},
  \href{http://dx.doi.org/10.1016/j.physrep.2023.01.003}{\emph{Phys. Rept.}
  {\bf 1010} (2023) 1--138}, [\href{http://arxiv.org/abs/2203.00019}{{\tt
  2203.00019}}].

\bibitem{Wistisen:2020czu}
T.~N. Wistisen, C.~H. Keitel and A.~Di~Piazza, \emph{{Transmutation of protons
  in a strong electromagnetic field}},
  \href{http://dx.doi.org/10.1088/1367-2630/abf705}{\emph{New J. Phys.} {\bf
  23} (2021) 065007}, [\href{http://arxiv.org/abs/2011.08031}{{\tt
  2011.08031}}].

\bibitem{Ouhammou:2022jys}
M.~Ouhammou, M.~Ouali, S.~Taj, R.~Benbrik and B.~Manaut, \emph{{Laser-induced
  proton decay}},
  \href{http://dx.doi.org/10.1007/s00340-023-08035-6}{\emph{Appl. Phys. B} {\bf
  129} (2023) 103}, [\href{http://arxiv.org/abs/2209.12191}{{\tt 2209.12191}}].

\bibitem{Ma:2025axq}
K.~Ma and T.~Li, \emph{{Laser induced Compton scattering to dark matter in
  effective field theory}},
  \href{http://dx.doi.org/10.1007/JHEP07(2025)028}{\emph{JHEP} {\bf 07} (2025)
  028}, [\href{http://arxiv.org/abs/2501.12687}{{\tt 2501.12687}}].

\bibitem{Fuchs:2024edo}
E.~Fuchs, F.~Kirk, E.~Madge, C.~Paranjape, E.~Peik, G.~Perez et~al.,
  \emph{{Implications of the laser excitation of the Th-229 nucleus for dark
  matter searches}},  \href{http://arxiv.org/abs/2407.15924}{{\tt 2407.15924}}.

\bibitem{Ma:2024ywm}
K.~Ma and T.~Li, \emph{{Laser induced Compton scattering to dark photon or
  axionlike particle}},
  \href{http://dx.doi.org/10.1103/PhysRevD.111.055001}{\emph{Phys. Rev. D} {\bf
  111} (2025) 055001}, [\href{http://arxiv.org/abs/2410.17591}{{\tt
  2410.17591}}].

\bibitem{Dillon:2018ypt}
B.~M. Dillon and B.~King, \emph{{ALP production through non-linear Compton
  scattering in intense fields}},
  \href{http://dx.doi.org/10.1140/epjc/s10052-018-6207-0}{\emph{Eur. Phys. J.
  C} {\bf 78} (2018) 775}, [\href{http://arxiv.org/abs/1802.07498}{{\tt
  1802.07498}}].

\bibitem{King:2018qbq}
B.~King, \emph{{Electron-seeded ALP production and ALP decay in an oscillating
  electromagnetic field}},
  \href{http://dx.doi.org/10.1016/j.physletb.2018.06.016}{\emph{Phys. Lett. B}
  {\bf 782} (2018) 737--743}, [\href{http://arxiv.org/abs/1802.07507}{{\tt
  1802.07507}}].

\bibitem{Bai:2021gbm}
Z.~Bai et~al., \emph{{New physics searches with an optical dump at LUXE}},
  \href{http://dx.doi.org/10.1103/PhysRevD.106.115034}{\emph{Phys. Rev. D} {\bf
  106} (2022) 115034}, [\href{http://arxiv.org/abs/2107.13554}{{\tt
  2107.13554}}].

\bibitem{Dillon:2018ouq}
B.~M. Dillon and B.~King, \emph{{Light scalars: coherent nonlinear Thomson
  scattering and detection}},
  \href{http://dx.doi.org/10.1103/PhysRevD.99.035048}{\emph{Phys. Rev. D} {\bf
  99} (2019) 035048}, [\href{http://arxiv.org/abs/1809.01356}{{\tt
  1809.01356}}].

\bibitem{King:2019cpj}
B.~King, B.~M. Dillon, K.~A. Beyer and G.~Gregori, \emph{{Axion-like-particle
  decay in strong electromagnetic backgrounds}},
  \href{http://dx.doi.org/10.1007/JHEP12(2019)162}{\emph{JHEP} {\bf 12} (2019)
  162}, [\href{http://arxiv.org/abs/1905.05201}{{\tt 1905.05201}}].

\bibitem{Beyer:2021mzq}
K.~A. Beyer, G.~Marocco, R.~Bingham and G.~Gregori,
  \emph{{Light-shining-through-wall axion detection experiments with a
  stimulating laser}},
  \href{http://dx.doi.org/10.1103/PhysRevD.105.035031}{\emph{Phys. Rev. D} {\bf
  105} (2022) 035031}, [\href{http://arxiv.org/abs/2109.14663}{{\tt
  2109.14663}}].

\bibitem{Huang:2020lxo}
S.~Huang, B.~Shen, Z.~Bu, X.~Zhang, L.~Ji and S.~Zhai, \emph{{Axion-like
  particle generation in laser-plasma interaction}},
  \href{http://dx.doi.org/10.1088/1402-4896/ac8b6b}{\emph{Phys. Scripta} {\bf
  97} (2022) 105303}, [\href{http://arxiv.org/abs/2005.02910}{{\tt
  2005.02910}}].

\bibitem{Born:1934gh}
M.~Born and L.~Infeld, \emph{{Foundations of the new field theory}},
  \href{http://dx.doi.org/10.1098/rspa.1934.0059}{\emph{Proc. Roy. Soc. Lond.
  A} {\bf 144} (1934) 425--451}.

\bibitem{Carley:2006zz}
H.~Carley and M.~K.~H. Kiessling, \emph{{Nonperturbative calculation of
  Born-Infeld effects on the Schroedinger spectrum of the hydrogen atom}},
  \href{http://dx.doi.org/10.1103/PhysRevLett.96.030402}{\emph{Phys. Rev.
  Lett.} {\bf 96} (2006) 030402},
  [\href{http://arxiv.org/abs/math-ph/0506069}{{\tt math-ph/0506069}}].

\bibitem{Davila:2013wba}
J.~M. Davila, C.~Schubert and M.~A. Trejo, \emph{{Photonic processes in
  Born-Infeld theory}},
  \href{http://dx.doi.org/10.1142/S0217751X14501747}{\emph{Int. J. Mod. Phys.
  A} {\bf 29} (2014) 1450174}, [\href{http://arxiv.org/abs/1310.8410}{{\tt
  1310.8410}}].

\bibitem{Fouche:2016qqj}
M.~Fouch{\'e}, R.~Battesti and C.~Rizzo, \emph{{Limits on nonlinear
  electrodynamics}},
  \href{http://dx.doi.org/10.1103/PhysRevD.93.093020}{\emph{Phys. Rev. D} {\bf
  93} (2016) 093020}, [\href{http://arxiv.org/abs/1605.04102}{{\tt
  1605.04102}}].

\bibitem{Ellis:2017edi}
J.~Ellis, N.~E. Mavromatos and T.~You, \emph{{Light-by-Light Scattering
  Constraint on Born-Infeld Theory}},
  \href{http://dx.doi.org/10.1103/PhysRevLett.118.261802}{\emph{Phys. Rev.
  Lett.} {\bf 118} (2017) 261802}, [\href{http://arxiv.org/abs/1703.08450}{{\tt
  1703.08450}}].

\bibitem{Peccei:1977hh}
R.~D. Peccei and H.~R. Quinn, \emph{{CP Conservation in the Presence of
  Instantons}},
  \href{http://dx.doi.org/10.1103/PhysRevLett.38.1440}{\emph{Phys. Rev. Lett.}
  {\bf 38} (1977) 1440--1443}.

\bibitem{Peccei:1977ur}
R.~D. Peccei and H.~R. Quinn, \emph{{Constraints Imposed by CP Conservation in
  the Presence of Instantons}},
  \href{http://dx.doi.org/10.1103/PhysRevD.16.1791}{\emph{Phys. Rev. D} {\bf
  16} (1977) 1791--1797}.

\bibitem{Weinberg:1977ma}
S.~Weinberg, \emph{{A New Light Boson?}},
  \href{http://dx.doi.org/10.1103/PhysRevLett.40.223}{\emph{Phys. Rev. Lett.}
  {\bf 40} (1978) 223--226}.

\bibitem{Wilczek:1977pj}
F.~Wilczek, \emph{{Problem of Strong $P$ and $T$ Invariance in the Presence of
  Instantons}}, \href{http://dx.doi.org/10.1103/PhysRevLett.40.279}{\emph{Phys.
  Rev. Lett.} {\bf 40} (1978) 279--282}.

\bibitem{DiLuzio:2020wdo}
L.~Di~Luzio, M.~Giannotti, E.~Nardi and L.~Visinelli, \emph{{The landscape of
  QCD axion models}},
  \href{http://dx.doi.org/10.1016/j.physrep.2020.06.002}{\emph{Phys. Rept.}
  {\bf 870} (2020) 1--117}, [\href{http://arxiv.org/abs/2003.01100}{{\tt
  2003.01100}}].

\bibitem{Dimopoulos:1979pp}
S.~Dimopoulos, \emph{{A Solution of the Strong {CP} Problem in Models With
  Scalars}}, \href{http://dx.doi.org/10.1016/0370-2693(79)91233-4}{\emph{Phys.
  Lett. B} {\bf 84} (1979) 435--439}.

\bibitem{Tye:1981zy}
S.~H.~H. Tye, \emph{{A Superstrong Force With a Heavy Axion}},
  \href{http://dx.doi.org/10.1103/PhysRevLett.47.1035}{\emph{Phys. Rev. Lett.}
  {\bf 47} (1981) 1035}.

\bibitem{Zhitnitsky:1980tq}
A.~R. Zhitnitsky, \emph{{On Possible Suppression of the Axion Hadron
  Interactions. (In Russian)}}, {\emph{Sov. J. Nucl. Phys.} {\bf 31} (1980)
  260}.

\bibitem{Dine:1981rt}
M.~Dine, W.~Fischler and M.~Srednicki, \emph{{A Simple Solution to the Strong
  CP Problem with a Harmless Axion}},
  \href{http://dx.doi.org/10.1016/0370-2693(81)90590-6}{\emph{Phys. Lett. B}
  {\bf 104} (1981) 199--202}.

\bibitem{Holdom:1982ex}
B.~Holdom and M.~E. Peskin, \emph{{Raising the Axion Mass}},
  \href{http://dx.doi.org/10.1016/0550-3213(82)90228-0}{\emph{Nucl. Phys. B}
  {\bf 208} (1982) 397--412}.

\bibitem{Kaplan:1985dv}
D.~B. Kaplan, \emph{{Opening the Axion Window}},
  \href{http://dx.doi.org/10.1016/0550-3213(85)90319-0}{\emph{Nucl. Phys. B}
  {\bf 260} (1985) 215--226}.

\bibitem{Srednicki:1985xd}
M.~Srednicki, \emph{{Axion Couplings to Matter. 1. CP Conserving Parts}},
  \href{http://dx.doi.org/10.1016/0550-3213(85)90054-9}{\emph{Nucl. Phys. B}
  {\bf 260} (1985) 689--700}.

\bibitem{Flynn:1987rs}
J.~M. Flynn and L.~Randall, \emph{{A Computation of the Small Instanton
  Contribution to the Axion Potential}},
  \href{http://dx.doi.org/10.1016/0550-3213(87)90089-7}{\emph{Nucl. Phys. B}
  {\bf 293} (1987) 731--739}.

\bibitem{Kamionkowski:1992mf}
M.~Kamionkowski and J.~March-Russell, \emph{{Planck scale physics and the
  Peccei-Quinn mechanism}},
  \href{http://dx.doi.org/10.1016/0370-2693(92)90492-M}{\emph{Phys. Lett. B}
  {\bf 282} (1992) 137--141}, [\href{http://arxiv.org/abs/hep-th/9202003}{{\tt
  hep-th/9202003}}].

\bibitem{Berezhiani:2000gh}
Z.~Berezhiani, L.~Gianfagna and M.~Giannotti, \emph{{Strong CP problem and
  mirror world: The Weinberg-Wilczek axion revisited}},
  \href{http://dx.doi.org/10.1016/S0370-2693(00)01392-7}{\emph{Phys. Lett. B}
  {\bf 500} (2001) 286--296}, [\href{http://arxiv.org/abs/hep-ph/0009290}{{\tt
  hep-ph/0009290}}].

\bibitem{Hsu:2004mf}
S.~D.~H. Hsu and F.~Sannino, \emph{{New solutions to the strong CP problem}},
  \href{http://dx.doi.org/10.1016/j.physletb.2004.11.040}{\emph{Phys. Lett. B}
  {\bf 605} (2005) 369--375}, [\href{http://arxiv.org/abs/hep-ph/0408319}{{\tt
  hep-ph/0408319}}].

\bibitem{Hook:2014cda}
A.~Hook, \emph{{Anomalous solutions to the strong CP problem}},
  \href{http://dx.doi.org/10.1103/PhysRevLett.114.141801}{\emph{Phys. Rev.
  Lett.} {\bf 114} (2015) 141801}, [\href{http://arxiv.org/abs/1411.3325}{{\tt
  1411.3325}}].

\bibitem{Alonso-Alvarez:2018irt}
G.~Alonso-\'Alvarez, M.~B. Gavela and P.~Quilez, \emph{{Axion couplings to
  electroweak gauge bosons}},
  \href{http://dx.doi.org/10.1140/epjc/s10052-019-6732-5}{\emph{Eur. Phys. J.
  C} {\bf 79} (2019) 223}, [\href{http://arxiv.org/abs/1811.05466}{{\tt
  1811.05466}}].

\bibitem{Hook:2019qoh}
A.~Hook, S.~Kumar, Z.~Liu and R.~Sundrum, \emph{{High Quality QCD Axion and the
  LHC}}, \href{http://dx.doi.org/10.1103/PhysRevLett.124.221801}{\emph{Phys.
  Rev. Lett.} {\bf 124} (2020) 221801},
  [\href{http://arxiv.org/abs/1911.12364}{{\tt 1911.12364}}].

\bibitem{Kim:1979if}
J.~E. Kim, \emph{{Weak Interaction Singlet and Strong CP Invariance}},
  \href{http://dx.doi.org/10.1103/PhysRevLett.43.103}{\emph{Phys. Rev. Lett.}
  {\bf 43} (1979) 103}.

\bibitem{Shifman:1979if}
M.~A. Shifman, A.~I. Vainshtein and V.~I. Zakharov, \emph{{Can Confinement
  Ensure Natural CP Invariance of Strong Interactions?}},
  \href{http://dx.doi.org/10.1016/0550-3213(80)90209-6}{\emph{Nucl. Phys. B}
  {\bf 166} (1980) 493--506}.

\bibitem{Turner:1989vc}
M.~S. Turner, \emph{{Windows on the Axion}},
  \href{http://dx.doi.org/10.1016/0370-1573(90)90172-X}{\emph{Phys. Rept.} {\bf
  197} (1990) 67--97}.

\bibitem{Rubakov:1997vp}
V.~A. Rubakov, \emph{{Grand unification and heavy axion}},
  \href{http://dx.doi.org/10.1134/1.567390}{\emph{JETP Lett.} {\bf 65} (1997)
  621--624}, [\href{http://arxiv.org/abs/hep-ph/9703409}{{\tt
  hep-ph/9703409}}].

\bibitem{Fukuda:2015ana}
H.~Fukuda, K.~Harigaya, M.~Ibe and T.~T. Yanagida, \emph{{Model of visible QCD
  axion}}, \href{http://dx.doi.org/10.1103/PhysRevD.92.015021}{\emph{Phys. Rev.
  D} {\bf 92} (2015) 015021}, [\href{http://arxiv.org/abs/1504.06084}{{\tt
  1504.06084}}].

\bibitem{Gherghetta:2016fhp}
T.~Gherghetta, N.~Nagata and M.~Shifman, \emph{{A Visible QCD Axion from an
  Enlarged Color Group}},
  \href{http://dx.doi.org/10.1103/PhysRevD.93.115010}{\emph{Phys. Rev. D} {\bf
  93} (2016) 115010}, [\href{http://arxiv.org/abs/1604.01127}{{\tt
  1604.01127}}].

\bibitem{Dimopoulos:2016lvn}
S.~Dimopoulos, A.~Hook, J.~Huang and G.~Marques-Tavares, \emph{{A collider
  observable QCD axion}},
  \href{http://dx.doi.org/10.1007/JHEP11(2016)052}{\emph{JHEP} {\bf 11} (2016)
  052}, [\href{http://arxiv.org/abs/1606.03097}{{\tt 1606.03097}}].

\bibitem{Chiang:2016eav}
C.-W. Chiang, H.~Fukuda, M.~Ibe and T.~T. Yanagida, \emph{{750 GeV diphoton
  resonance in a visible heavy QCD axion model}},
  \href{http://dx.doi.org/10.1103/PhysRevD.93.095016}{\emph{Phys. Rev. D} {\bf
  93} (2016) 095016}, [\href{http://arxiv.org/abs/1602.07909}{{\tt
  1602.07909}}].

\bibitem{Gaillard:2018xgk}
M.~K. Gaillard, M.~B. Gavela, R.~Houtz, P.~Quilez and R.~Del~Rey, \emph{{Color
  unified dynamical axion}},
  \href{http://dx.doi.org/10.1140/epjc/s10052-018-6396-6}{\emph{Eur. Phys. J.
  C} {\bf 78} (2018) 972}, [\href{http://arxiv.org/abs/1805.06465}{{\tt
  1805.06465}}].

\bibitem{Gherghetta:2020ofz}
T.~Gherghetta and M.~D. Nguyen, \emph{{A Composite Higgs with a Heavy Composite
  Axion}}, \href{http://dx.doi.org/10.1007/JHEP12(2020)094}{\emph{JHEP} {\bf
  12} (2020) 094}, [\href{http://arxiv.org/abs/2007.10875}{{\tt 2007.10875}}].

\bibitem{Knapen:2016moh}
S.~Knapen, T.~Lin, H.~K. Lou and T.~Melia, \emph{{Searching for Axionlike
  Particles with Ultraperipheral Heavy-Ion Collisions}},
  \href{http://dx.doi.org/10.1103/PhysRevLett.118.171801}{\emph{Phys. Rev.
  Lett.} {\bf 118} (2017) 171801}, [\href{http://arxiv.org/abs/1607.06083}{{\tt
  1607.06083}}].

\bibitem{Baldenegro:2018hng}
C.~Baldenegro, S.~Fichet, G.~von Gersdorff and C.~Royon, \emph{{Searching for
  axion-like particles with proton tagging at the LHC}},
  \href{http://dx.doi.org/10.1007/JHEP06(2018)131}{\emph{JHEP} {\bf 06} (2018)
  131}, [\href{http://arxiv.org/abs/1803.10835}{{\tt 1803.10835}}].

\bibitem{Inan:2020aal}
S.~C. \.Inan and A.~V. Kisselev, \emph{{A search for axion-like particles in
  light-by-light scattering at the CLIC}},
  \href{http://dx.doi.org/10.1007/JHEP06(2020)183}{\emph{JHEP} {\bf 06} (2020)
  183}, [\href{http://arxiv.org/abs/2003.01978}{{\tt 2003.01978}}].

\bibitem{Inan:2020kif}
S.~C. \.Inan and A.~V. Kisselev, \emph{{Polarized light-by-light scattering at
  the CLIC induced by axion-like particles}},
  \href{http://dx.doi.org/10.1088/1674-1137/abe0be}{\emph{Chin. Phys. C} {\bf
  45} (2021) 043109}, [\href{http://arxiv.org/abs/2007.01693}{{\tt
  2007.01693}}].

\bibitem{Harland-Lang:2022jwn}
L.~A. Harland-Lang and M.~Tasevsky, \emph{{New calculation of semiexclusive
  axionlike particle production at the LHC}},
  \href{http://dx.doi.org/10.1103/PhysRevD.107.033001}{\emph{Phys. Rev. D} {\bf
  107} (2023) 033001}, [\href{http://arxiv.org/abs/2208.10526}{{\tt
  2208.10526}}].

\bibitem{Balkin:2023gya}
R.~Balkin, O.~Hen, W.~Li, H.~Liu, T.~Ma, Y.~Soreq et~al., \emph{{Probing
  axion-like particles at the Electron-Ion Collider}},
  \href{http://dx.doi.org/10.1007/JHEP02(2024)123}{\emph{JHEP} {\bf 02} (2024)
  123}, [\href{http://arxiv.org/abs/2310.08827}{{\tt 2310.08827}}].

\bibitem{RebelloTeles:2023uig}
P.~Rebello~Teles, D.~d'Enterria, V.~P. Gon{\c{c}}alves and D.~E. Martins,
  \emph{{Searches for axionlike particles via
  {\ensuremath{\gamma}}{\ensuremath{\gamma}} fusion at future e+e- colliders}},
  \href{http://dx.doi.org/10.1103/PhysRevD.109.055003}{\emph{Phys. Rev. D} {\bf
  109} (2024) 055003}, [\href{http://arxiv.org/abs/2310.17270}{{\tt
  2310.17270}}].

\bibitem{ATLAS:2023zfc}
{\scshape ATLAS} collaboration, G.~Aad et~al., \emph{{Search for an axion-like
  particle with forward proton scattering in association with photon pairs at
  ATLAS}}, \href{http://dx.doi.org/10.1007/JHEP07(2023)234}{\emph{JHEP} {\bf
  07} (2023) 234}, [\href{http://arxiv.org/abs/2304.10953}{{\tt 2304.10953}}].

\bibitem{TOTEM:2021zxa}
{\scshape TOTEM, CMS} collaboration, A.~Tumasyan et~al., \emph{{First Search
  for Exclusive Diphoton Production at High Mass with Tagged Protons in
  Proton-Proton Collisions at $\sqrt s$ = 13 TeV}},
  \href{http://dx.doi.org/10.1103/PhysRevLett.129.011801}{\emph{Phys. Rev.
  Lett.} {\bf 129} (2022) 011801}, [\href{http://arxiv.org/abs/2110.05916}{{\tt
  2110.05916}}].

\bibitem{TOTEM:2023ewz}
{\scshape TOTEM, CMS} collaboration, A.~Tumasyan et~al., \emph{{Search for
  high-mass exclusive diphoton production with tagged protons in proton-proton
  collisions at s=13{\,}{\,}TeV}},
  \href{http://dx.doi.org/10.1103/PhysRevD.110.012010}{\emph{Phys. Rev. D} {\bf
  110} (2024) 012010}, [\href{http://arxiv.org/abs/2311.02725}{{\tt
  2311.02725}}].

\bibitem{Heinzl:2024cia}
T.~Heinzl, B.~King and D.~Liu, \emph{{Coherent enhancement of QED cross
  sections in electromagnetic backgrounds}},
  \href{http://dx.doi.org/10.1103/PhysRevD.111.056018}{\emph{Phys. Rev. D} {\bf
  111} (2025) 056018}, [\href{http://arxiv.org/abs/2412.10574}{{\tt
  2412.10574}}].

\bibitem{Jikia:1993tc}
G.~Jikia and A.~Tkabladze, \emph{{Photon-photon scattering at the photon linear
  collider}}, \href{http://dx.doi.org/10.1016/0370-2693(94)91246-7}{\emph{Phys.
  Lett. B} {\bf 323} (1994) 453--458},
  [\href{http://arxiv.org/abs/hep-ph/9312228}{{\tt hep-ph/9312228}}].

\bibitem{Gounaris:1998qk}
G.~J. Gounaris, P.~I. Porfyriadis and F.~M. Renard, \emph{{Light by light
  scattering at high-energy: A Tool to reveal new particles}},
  \href{http://dx.doi.org/10.1016/S0370-2693(99)00171-9}{\emph{Phys. Lett. B}
  {\bf 452} (1999) 76--82}, [\href{http://arxiv.org/abs/hep-ph/9812378}{{\tt
  hep-ph/9812378}}].

\bibitem{Gounaris:1999gh}
G.~J. Gounaris, P.~I. Porfyriadis and F.~M. Renard, \emph{{The gamma gamma
  ---\ensuremath{>} gamma gamma process in the standard and SUSY models at
  high-energies}}, \href{http://dx.doi.org/10.1007/s100529900079}{\emph{Eur.
  Phys. J. C} {\bf 9} (1999) 673--686},
  [\href{http://arxiv.org/abs/hep-ph/9902230}{{\tt hep-ph/9902230}}].

\bibitem{Bern:2001dg}
Z.~Bern, A.~De~Freitas, L.~J. Dixon, A.~Ghinculov and H.~L. Wong, \emph{{QCD
  and QED corrections to light by light scattering}},
  \href{http://dx.doi.org/10.1088/1126-6708/2001/11/031}{\emph{JHEP} {\bf 11}
  (2001) 031}, [\href{http://arxiv.org/abs/hep-ph/0109079}{{\tt
  hep-ph/0109079}}].

\bibitem{Passarino:1978jh}
G.~Passarino and M.~J.~G. Veltman, \emph{{One Loop Corrections for e+ e-
  Annihilation Into mu+ mu- in the Weinberg Model}},
  \href{http://dx.doi.org/10.1016/0550-3213(79)90234-7}{\emph{Nucl. Phys. B}
  {\bf 160} (1979) 151--207}.

\bibitem{Yakimenko:2019sya}
V.~Yakimenko et~al., \emph{{FACET-II facility for advanced accelerator
  experimental tests}},
  \href{http://dx.doi.org/10.1103/PhysRevAccelBeams.22.101301}{\emph{Phys. Rev.
  Accel. Beams} {\bf 22} (2019) 101301}.

\bibitem{Altarelli:2006zza}
M.~Altarelli et~al., \emph{{XFEL: The European X-Ray Free-Electron Laser.
  Technical design report}}, .

\bibitem{Burke:1997ew}
D.~L. Burke et~al., \emph{{Positron production in multi - photon light by light
  scattering}},
  \href{http://dx.doi.org/10.1103/PhysRevLett.79.1626}{\emph{Phys. Rev. Lett.}
  {\bf 79} (1997) 1626--1629}.

\bibitem{Bamber:1999zt}
C.~Bamber et~al., \emph{{Studies of nonlinear QED in collisions of 46.6-GeV
  electrons with intense laser pulses}},
  \href{http://dx.doi.org/10.1103/PhysRevD.60.092004}{\emph{Phys. Rev. D} {\bf
  60} (1999) 092004}.

\bibitem{Ackermann:2007zzd}
W.~Ackermann et~al., \emph{{Operation of a free-electron laser from the extreme
  ultraviolet to the water window}},
  \href{http://dx.doi.org/10.1038/nphoton.2007.76}{\emph{Nature Photon.} {\bf
  1} (2007) 336--342}.

\bibitem{PhysRevX.7.021043}
P.~Finetti, H.~H\"oppner, E.~Allaria, C.~Callegari, F.~Capotondi,
  P.~Cinquegrana et~al., \emph{Pulse duration of seeded free-electron lasers},
  \href{http://dx.doi.org/10.1103/PhysRevX.7.021043}{\emph{Phys. Rev. X} {\bf
  7} (Jun, 2017) 021043}.

\bibitem{Rebhan:2017zdx}
A.~Rebhan and G.~Turk, \emph{{Polarization effects in light-by-light
  scattering: Euler\textendash{}Heisenberg versus Born\textendash{}Infeld}},
  \href{http://dx.doi.org/10.1142/S0217751X17500531}{\emph{Int. J. Mod. Phys.
  A} {\bf 32} (2017) 1750053}, [\href{http://arxiv.org/abs/1701.07375}{{\tt
  1701.07375}}].

\bibitem{Brivio:2017ije}
I.~Brivio, M.~B. Gavela, L.~Merlo, K.~Mimasu, J.~M. No, R.~del Rey et~al.,
  \emph{{ALPs Effective Field Theory and Collider Signatures}},
  \href{http://dx.doi.org/10.1140/epjc/s10052-017-5111-3}{\emph{Eur. Phys. J.
  C} {\bf 77} (2017) 572}, [\href{http://arxiv.org/abs/1701.05379}{{\tt
  1701.05379}}].

\bibitem{Herr:2003em}
W.~Herr and B.~Muratori, \emph{{Concept of luminosity}},  in \emph{{CERN
  Accelerator School and DESY Zeuthen: Accelerator Physics}}, pp.~361--377, 9,
  2003.

\bibitem{Cannoni:2016hro}
M.~Cannoni, \emph{{Lorentz invariant relative velocity and relativistic binary
  collisions}}, \href{http://dx.doi.org/10.1142/S0217751X17300022}{\emph{Int.
  J. Mod. Phys. A} {\bf 32} (2017) 1730002},
  [\href{http://arxiv.org/abs/1605.00569}{{\tt 1605.00569}}].

\bibitem{LUXE:2023crk}
{\scshape LUXE} collaboration, H.~Abramowicz et~al., \emph{{Technical Design
  Report for the LUXE Experiment}},
  \href{http://arxiv.org/abs/2308.00515}{{\tt 2308.00515}}.

\bibitem{Sarri:2025qng}
G.~Sarri et~al., \emph{{Input to the European Strategy for Particle Physics:
  Strong-Field Quantum Electrodynamics}},
  \href{http://arxiv.org/abs/2504.02608}{{\tt 2504.02608}}.

\bibitem{Jaeckel:2015jla}
J.~Jaeckel and M.~Spannowsky, \emph{{Probing MeV to 90 GeV axion-like particles
  with LEP and LHC}},
  \href{http://dx.doi.org/10.1016/j.physletb.2015.12.037}{\emph{Phys. Lett. B}
  {\bf 753} (2016) 482--487}, [\href{http://arxiv.org/abs/1509.00476}{{\tt
  1509.00476}}].

\bibitem{BESIII:2022rzz}
{\scshape BESIII} collaboration, M.~Ablikim et~al., \emph{{Search for an
  axion-like particle in radiative J/{\ensuremath{\psi}} decays}},
  \href{http://dx.doi.org/10.1016/j.physletb.2023.137698}{\emph{Phys. Lett. B}
  {\bf 838} (2023) 137698}, [\href{http://arxiv.org/abs/2211.12699}{{\tt
  2211.12699}}].

\bibitem{BESIII:2024hdv}
{\scshape BESIII} collaboration, M.~Ablikim et~al., \emph{{Search for diphoton
  decays of an axionlike particle in radiative J/{\ensuremath{\psi}} decays}},
  \href{http://dx.doi.org/10.1103/PhysRevD.110.L031101}{\emph{Phys. Rev. D}
  {\bf 110} (2024) L031101}, [\href{http://arxiv.org/abs/2404.04640}{{\tt
  2404.04640}}].

\bibitem{Belle-II:2020jti}
{\scshape Belle-II} collaboration, F.~Abudin{\'e}n et~al., \emph{{Search for
  Axion-Like Particles produced in $e^+e^-$ collisions at Belle II}},
  \href{http://dx.doi.org/10.1103/PhysRevLett.125.161806}{\emph{Phys. Rev.
  Lett.} {\bf 125} (2020) 161806}, [\href{http://arxiv.org/abs/2007.13071}{{\tt
  2007.13071}}].

\bibitem{BERGSMA1985458}
F.~Bergsma, J.~Dorenbosch, J.~Allaby, U.~Amaldi, G.~Barbiellini, C.~Berger
  et~al., \emph{Search for axion-like particle production in 400 gev
  proton-copper interactions},
  \href{http://dx.doi.org/https://doi.org/10.1016/0370-2693(85)90400-9}{\emph{Physics
  Letters B} {\bf 157} (1985) 458--462}.

\bibitem{PhysRevLett.59.755}
E.~M. Riordan, M.~W. Krasny, K.~Lang, P.~de~Barbaro, A.~Bodek, S.~Dasu et~al.,
  \emph{Search for short-lived axions in an electron-beam-dump experiment},
  \href{http://dx.doi.org/10.1103/PhysRevLett.59.755}{\emph{Phys. Rev. Lett.}
  {\bf 59} (Aug, 1987) 755--758}.

\bibitem{Dolan:2017osp}
M.~J. Dolan, T.~Ferber, C.~Hearty, F.~Kahlhoefer and K.~Schmidt-Hoberg,
  \emph{{Revised constraints and Belle II sensitivity for visible and invisible
  axion-like particles}},
  \href{http://dx.doi.org/10.1007/JHEP12(2017)094}{\emph{JHEP} {\bf 12} (2017)
  094}, [\href{http://arxiv.org/abs/1709.00009}{{\tt 1709.00009}}].

\bibitem{PhysRevD.38.3375}
J.~D. Bjorken, S.~Ecklund, W.~R. Nelson, A.~Abashian, C.~Church, B.~Lu et~al.,
  \emph{Search for neutral metastable penetrating particles produced in the
  slac beam dump},
  \href{http://dx.doi.org/10.1103/PhysRevD.38.3375}{\emph{Phys. Rev. D} {\bf
  38} (Dec, 1988) 3375--3386}.

\bibitem{Dobrich:2019dxc}
B.~D{\"o}brich, J.~Jaeckel and T.~Spadaro, \emph{{Light in the beam dump - ALP
  production from decay photons in proton beam-dumps}},
  \href{http://dx.doi.org/10.1007/JHEP05(2019)213}{\emph{JHEP} {\bf 05} (2019)
  213}, [\href{http://arxiv.org/abs/1904.02091}{{\tt 1904.02091}}].

\bibitem{Blumlein:1990ay}
J.~Blumlein et~al., \emph{{Limits on neutral light scalar and pseudoscalar
  particles in a proton beam dump experiment}},
  \href{http://dx.doi.org/10.1007/BF01548556}{\emph{Z. Phys. C} {\bf 51} (1991)
  341--350}.

\bibitem{NA64:2020qwq}
{\scshape NA64} collaboration, D.~Banerjee et~al., \emph{{Search for Axionlike
  and Scalar Particles with the NA64 Experiment}},
  \href{http://dx.doi.org/10.1103/PhysRevLett.125.081801}{\emph{Phys. Rev.
  Lett.} {\bf 125} (2020) 081801}, [\href{http://arxiv.org/abs/2005.02710}{{\tt
  2005.02710}}].

\bibitem{dEnterria:2021ljz}
D.~d'Enterria, \emph{{Collider constraints on axion-like particles}},  in
  \emph{{Workshop on Feebly Interacting Particles}}, 2, 2021.
\newblock \href{http://arxiv.org/abs/2102.08971}{{\tt 2102.08971}}.

\bibitem{BaBar:2017tiz}
{\scshape BaBar} collaboration, J.~P. Lees et~al., \emph{{Search for Invisible
  Decays of a Dark Photon Produced in ${e}^{+}{e}^{-}$ Collisions at BaBar}},
  \href{http://dx.doi.org/10.1103/PhysRevLett.119.131804}{\emph{Phys. Rev.
  Lett.} {\bf 119} (2017) 131804}, [\href{http://arxiv.org/abs/1702.03327}{{\tt
  1702.03327}}].

\end{thebibliography}\endgroup

\end{document}